\pdfoutput=1
\documentclass[11pt,twoside,a4paper,cmspaper,final,collab]{cms-tdr}
\def\svnVersion{541bfed}\def\svnDate{2024/05/15}\def\cmsCernNoTag{CERN-EP-2023-213}\def\cmsCernDate{\today}\def\cmsMessage{Published in the Journal of High Energy Physics as \href{http://dx.doi.org/10.1007/JHEP05(2024)046}{\doi{10.1007/JHEP05(2024)046}.}}
\begin{document}\cmsNoteHeader{B2G-20-012}

\newcommand{\wjets}{\ensuremath{\PW+\text{jets}}\xspace}
\newcommand{\DeepJet}{\textsc{DeepJet}\xspace}
\newcommand{\abseta}{\ensuremath{\abs{\eta}}\xspace}
\newcommand{\jetwprime}{\ensuremath{\mathrm{j_{\PWpr}}}\xspace}
\newcommand{\jettop}{\ensuremath{\mathrm{j_{\PQt}}}\xspace}
\newcommand{\jetAKeight}{\ensuremath{\mathrm{j_{AK8}}}\xspace}
\newcommand{\mtop}{\ensuremath{M_{\PQt}}\xspace}
\newcommand{\msd}{\ensuremath{M_{\mathrm{SD,AK8}}}\xspace}
\newcommand{\mwp}{\ensuremath{m_{\PWpr}}\xspace}
\newcommand{\pznu}{\ensuremath{p_{z}^{\PGn}}}
\newcommand{\mlvjj}{\ensuremath{M_{\mathrm{\Pell \PGn jj}}}\xspace}
\newcommand{\PFminiso}[1]{\ensuremath{I_{\text{mini}}^{\mathrm{#1}}}\xspace}

\newcommand{\rz}{\ensuremath{\mathrm{R0}}\xspace}
\newcommand{\rtwob}{\ensuremath{\mathrm{R2B}}\xspace}
\newcommand{\rw}{\ensuremath{\mathrm{R\PWpr}}\xspace}
\newcommand{\rt}{\ensuremath{\mathrm{RT}}\xspace}

\newcommand{\rza}{\ensuremath{\mathrm{\rz_A}}\xspace}

\newcommand{\rtwoba}{\ensuremath{\mathrm{\rtwob_A}}\xspace}

\newcommand{\rwa}{\ensuremath{\mathrm{\rw_A}}\xspace}

\newcommand{\rta}{\ensuremath{\mathrm{\rt_A}}\xspace}

\newcommand{\rbig}{\ensuremath{\mathrm{R}}\xspace}
\newcommand{\ra}{\ensuremath{\mathrm{R_A}}\xspace}
\newcommand{\rb}{\ensuremath{\mathrm{R_B}}\xspace}
\newcommand{\rc}{\ensuremath{\mathrm{R_C}}\xspace}
\newcommand{\rdd}{\ensuremath{\mathrm{R_D}}\xspace}

\newcommand{\relwidth}{\ensuremath{\Gamma/m_{\PWpr}}\xspace}

\newcommand{\TF}{\ensuremath{TF (\mlvjj)}\xspace}
\newcommand{\TFbin}{\ensuremath{\text{TF}}\xspace}
\newcommand{\CF}{\ensuremath{\text{CF}}\xspace}
\newcommand{\cmsTable}[1]{\resizebox{\textwidth}{!}{#1}}

\cmsNoteHeader{B2G-20-012} 
\title{Search for \texorpdfstring{\PWpr}{W'} bosons decaying to a top and a bottom quark in leptonic final states in proton-proton collisions at \texorpdfstring{$\sqrt{s} = 13\TeV$}{sqrt(s) = 13 TeV}}
\date{\today}

\abstract{A search for \PWpr bosons decaying to a top and a bottom quark in final states including an electron or a muon is performed with the CMS detector at the LHC. The analyzed data correspond to an integrated luminosity of 138\fbinv of proton-proton collisions at a center-of-mass energy of 13\TeV. Good agreement with the standard model expectation is observed and no evidence for the existence of the \PWpr boson is found over the mass range examined. The largest observed deviation from the standard model expectation is found for a \PWpr boson mass ($m_{\PWpr}$) hypothesis of 3.8\TeV with a relative decay width of 1\%, with a local (global) significance of 2.6 (2.0) standard deviations. Upper limits on the production cross sections of \PWpr bosons decaying to a top and a bottom quark are set. Left- and right-handed \PWpr bosons with $m_{\PWpr}$ below 3.9 and 4.3\TeV, respectively, are excluded at the 95\% confidence level, under the assumption that the new particle has a narrow decay width.  Limits are also set for relative decay widths up to 30\%.}

\hypersetup{
pdfauthor={CMS Collaboration},
pdftitle={Search for W' bosons decaying to a top and a bottom quark in leptonic final states in proton-proton collisions at sqrt(s) = 13 TeV},
pdfsubject={CMS},
pdfkeywords={CMS, W' search, top resonances, boosted top}}

\maketitle

\section{Introduction}
\label{sec:intro}
Despite the remarkable success of the standard model (SM), it is not a complete theory since it is not able to explain several experimental observations.
Among the notable phenomena that may require the existence of physics beyond the SM are the apparent existence of dark matter, the matter-antimatter asymmetry in the universe, and neutrino oscillations. The SM also does not provide an explanation for the origin of its internal structure, such as the hierarchy of fermion masses, and suffers from the fine-tuning problem, which reflects the need to keep quantum corrections to the Higgs boson mass finite. All of these phenomena suggest the existence of a more fundamental theory.

Many models have emerged to provide an explanation for these open issues, often introducing new particles in the energy range reachable at the CERN Large Hadron Collider (LHC). Several of these models predict the existence of new massive bosons, called \PWpr and \PZpr, with similar properties to their SM counterparts, the \PW and \PZ bosons, respectively, that act as mediators of the electroweak interaction~\cite{PhysRevD.11.2558, Senjanovic:1975rk,Minkowski:1977sc,Mohapatra:1979ia, Pati:1980rx, Mohapatra:1986uf, PhysRevD.36.878, PhysRevD.46.410, PhysRevD.74.075008, Chacko:2005un, PhysRevD.76.115018, Belyaev:2008yj, Cabarcas:2009vb, Buras:2012dp, Ahmed:2017psb}.

Models with \PWpr and \PZpr bosons, coupled preferentially to third-generation fermions, are of particular interest as they could be involved in the explanation of flavor anomalies~\cite{BaBar:2012obs, Belle:2016dyj, PhysRevD.98.055016}, or in the mechanism of electroweak symmetry breaking~\cite{Muller:1996dj,PhysRevD.104.055006}. 
In this context, the top quark plays a particularly important role, as it has both a large Yukawa coupling to the Higgs boson and a distinctive experimental signature due to its decay chain.
Dedicated searches for \PWpr and \PZpr bosons coupled preferentially to third-generation quarks and leptons have been performed in the past by both the CMS and ATLAS Collaborations~\cite{Sirunyan:2017vkm_,ATLAS:2018ihk,ATLAS:2018wmg,Sirunyan_2019_,CMS:2021mux_, ATLAS:2023ibb}. No significant evidence for the existence of these particles has been found.

\begin{figure*}[hbtp]
\begin{center}
\includegraphics[width=0.6\textwidth]{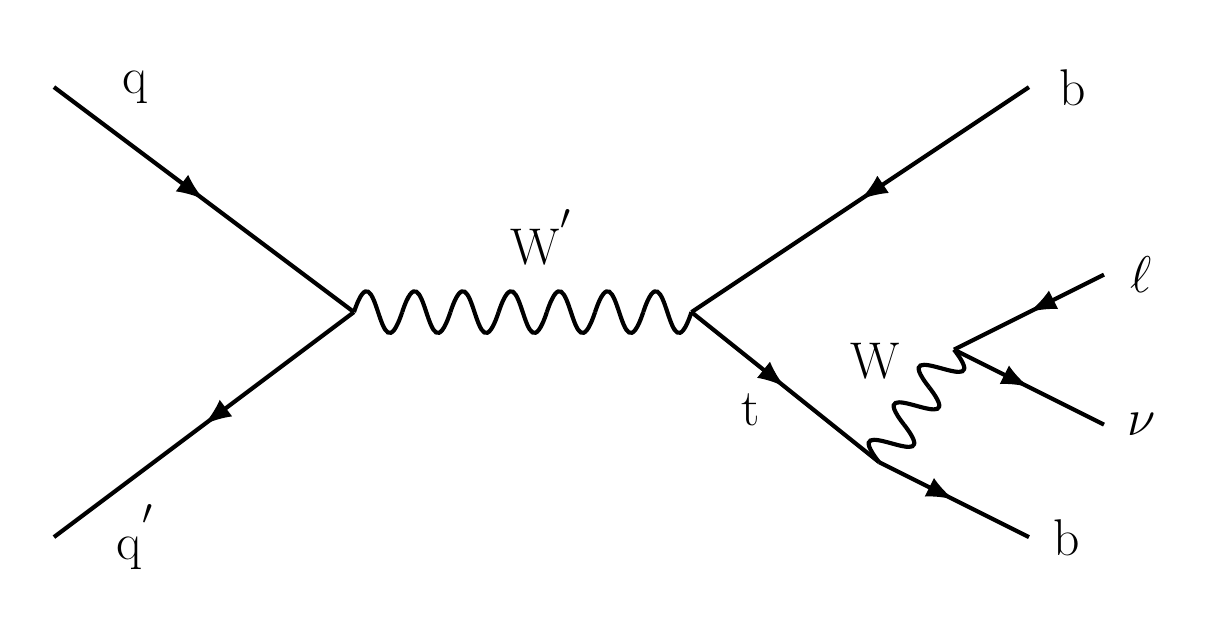}
\end{center}
\caption{Representative LO Feynman diagram for a \PWpr boson produced in the $s$-channel and decaying to a top and a bottom quark, with a lepton in the final state.}
\label{fig:feyndiagram}
\end{figure*}

In this paper we present a search for a \PWpr boson decaying to a top and a bottom quark, targeting the high end of the accessible mass spectrum at the LHC in the multi-\TeVns range.
The analysis focuses on the decay chains of the hypothesized \PWpr boson including leptons, \ie, $\PWpr \to \PQt\PQb \to \PW \PQb\PQb \to \ell \PGn \PQb\PQb$, where $\ell$ represents either an electron or a muon, including the leptonic decays of the \PGt lepton.
A representative leading-order (LO) Feynman diagram of the process considered here is shown in Fig.~\ref{fig:feyndiagram}.

Hypotheses for the new particle mass, \mwp, are considered in the range 2-6\TeV, and the analysis strategy is tailored towards reconstructing highly energetic top quarks with a lepton in the final state, complementing previous searches in the range below 3\TeV~\cite{CMS:2021mux_,Sirunyan:2017vkm_}. Different hypotheses for the width as well as the chirality of the new particle are explored to allow for an interpretation in a wide array of models.

Depending on the model, the relative decay width of the \PWpr boson with respect to its mass (\relwidth) could be significant~\cite{Belyaev:2008yj,Vignaroli:2014bpa,Barducci:2015vyf,PhysRevD.104.055006}, resulting in signatures that could escape standard searches. This analysis probes \relwidth of 1, 10, 20, and 30$\%$, for the first time.
The total width of approximately 1\% (more precisely 0.8\%) corresponds to the case where the \PWpr boson is only allowed to decay into \PQt and \PQb quarks.
In the model used here, the couplings, and thus the partial widths, are kept fixed and not varied together with the total width. In this interpretation, the larger width can be attributed to the presence of additional decays, as several models predict \PWpr bosons decaying to additional new particles~\cite{Vignaroli:2014bpa,Barducci:2015vyf}. This leads to a reduced branching fraction compared to only considering $\PWpr \to \PQt\PQb$ decays.
In the case of fixed couplings, the cross section decreases with larger width as this implies a smaller braching fraction to $\PQt\PQb$ pairs. This is opposite to the case of a fixed branching fraction, in which the width increases because of a larger coupling strength, which leads to a larger cross section.
We consider cases where the chirality of the \PWpr boson is left-handed (LH), right-handed (RH), or a combination of the two. For the LH, the signal is simulated including the SM production of single top quarks in the $s$-channel and the relative interference with the production with the \PWpr boson, which leads to a different signal phenomenology, depending on the mixing of the chiral components.

\begin{figure}[hbtp]
\centering
\includegraphics[width=0.48\textwidth]{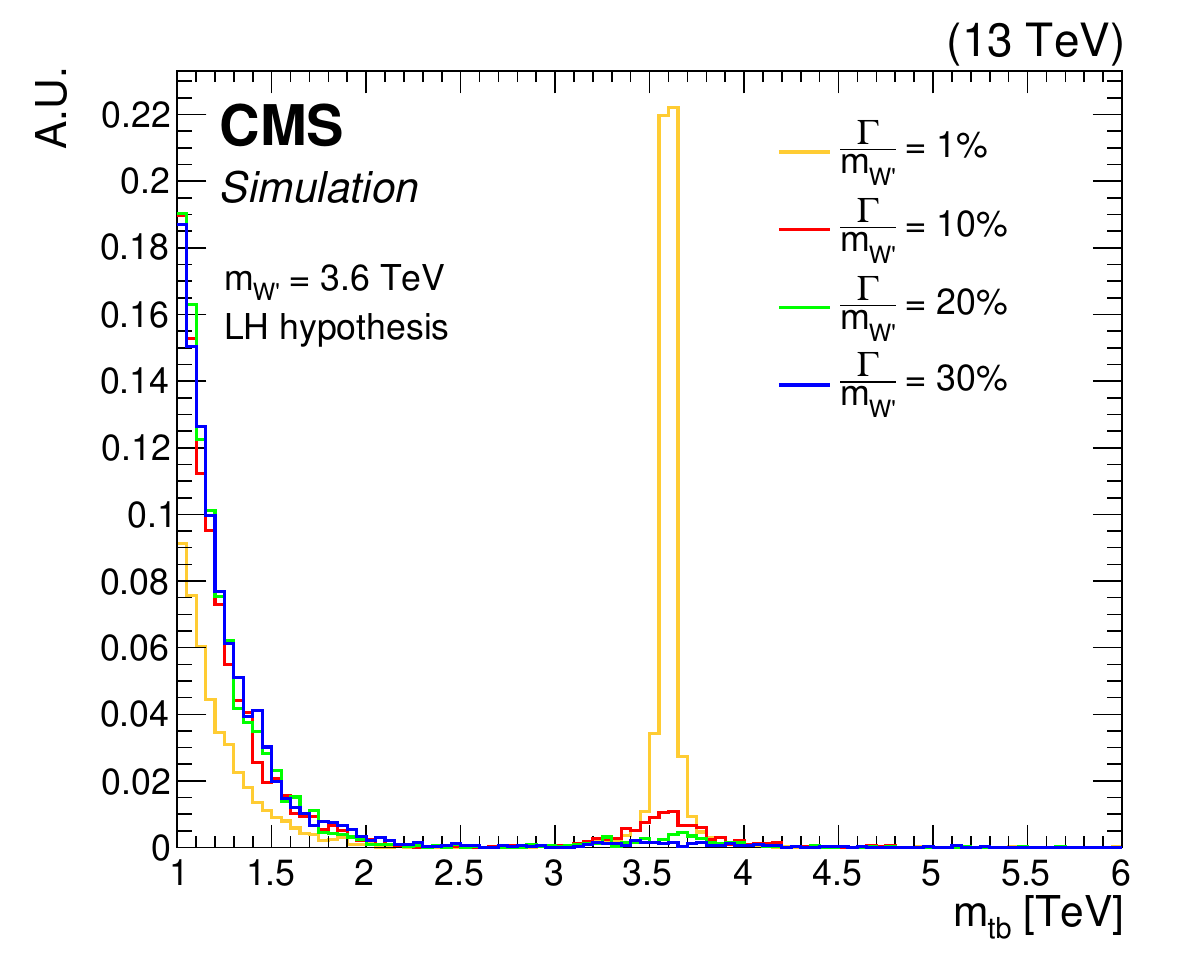}
\includegraphics[width=0.48\textwidth]{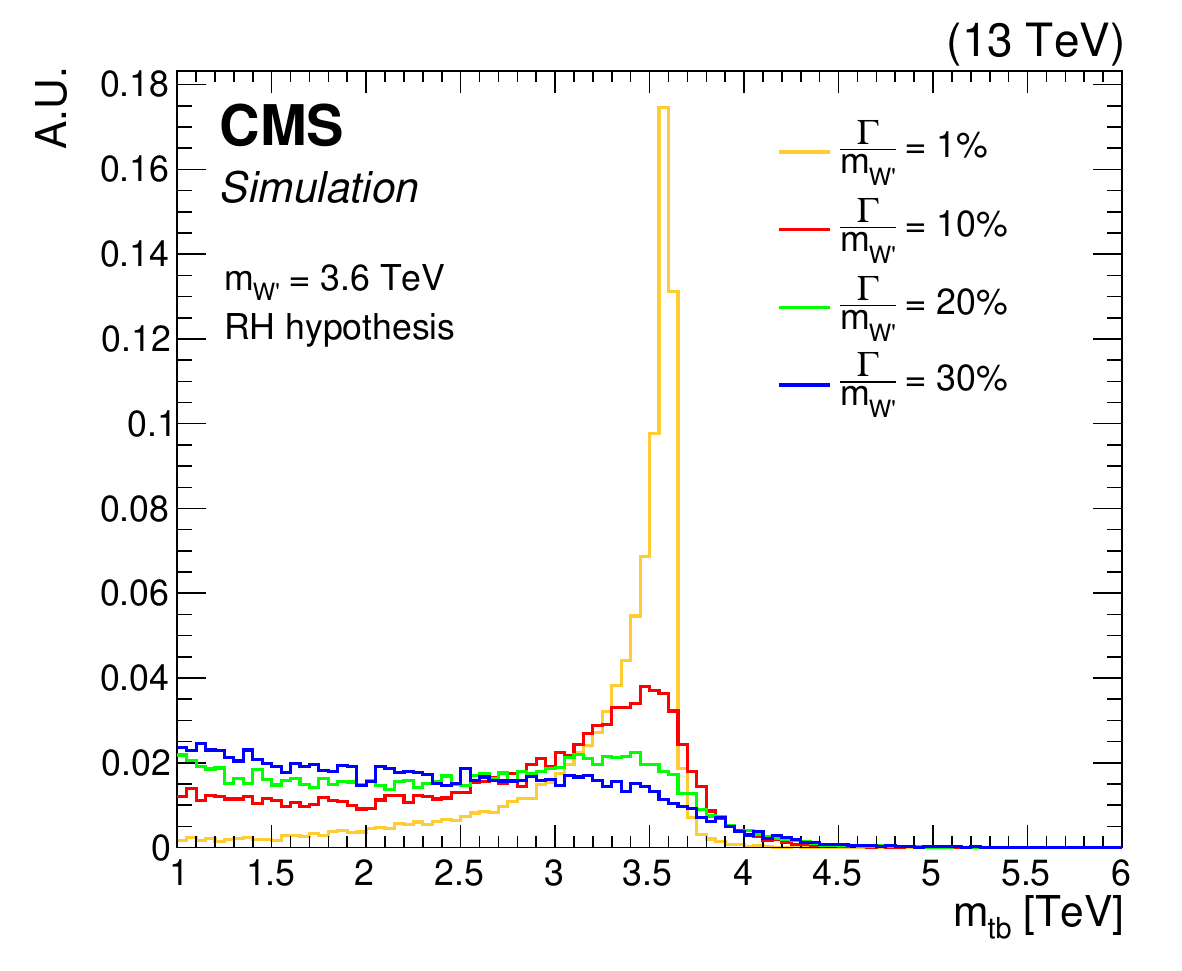}
\caption{Representative distributions of the invariant mass of the top-bottom quark pair, $m_{\PQt\PQb}$, as originating from the \PWpr boson for left- (left) and right-handed (right) \PWpr bosons, with relative widths \relwidth of 1, 10, 20, and 30$\%$ for a \PWpr boson mass of 3.6\TeV. For the LH, the signal is simulated including the SM production of single top quarks in the $s$-channel to correctly take into account the relative interference with the production of the \PWpr boson.}
\label{fig:particlewidths}
\end{figure}

Figure~\ref{fig:particlewidths} shows representative distributions of the mass of the system composed of the top-bottom quark pair as originating from the \PWpr boson for narrow- and large-width samples, for left- and right-handed chiralities, respectively, for a \PWpr boson mass of 3.6\TeV.

The width of the \PWpr boson affects the reconstructed mass distribution, resulting in a broader peak and an asymmetry favoring lower values.  This is quite visible for the cases with large decay width, where the tail towards small masses is dominant because of off-shell \PWpr boson production, enhanced by rapidly increasing parton distribution functions (PDFs) for decreasing partonic momentum fractions.
For the left-handed hypothesis, the region of the \mwp spectrum below 2\TeV is dominated by SM $s$-channel single top quark production, yielding a very different signal shape compared to the RH case.
 
This article is organized as follows: Section~\ref{sec:CMS} gives a brief description of the CMS apparatus, Sections~\ref{sec:samples} and~\ref{sec:samples_sim} report details on the data set and simulated samples, respectively, used in the analysis. Section~\ref{sec:event_sel} describes the requirements for the selection of events, while Sections~\ref{sec:reconstruction} and~\ref{sec:categorization} report the strategy for the reconstruction and the categorization of the events selected for analysis. The procedure for the prediction of most important backgrounds is described in Section~\ref{sec:bkgmodelling}, the systematic uncertainties affecting the analysis are reported in Section~\ref{sec:systematics}, and the limit extraction and results are detailed in Section~\ref{sec:results}. Tabulated results are provided in the HEPData record for this analysis~\cite{hepdata}.

\section{The CMS detector} 
\label{sec:CMS}
The central feature of the CMS apparatus is a superconducting solenoid of 6\unit{m} internal diameter, providing a magnetic field of 3.8\unit{T}. Within the solenoid volume are a silicon pixel and silicon strip tracker, a lead tungstate crystal electromagnetic calorimeter (ECAL), and a brass-and-scintillator hadron calorimeter (HCAL), each composed of a barrel and two endcap sections. Forward calorimeters extend the coverage in pseudorapidity ($\eta$) provided by the barrel and endcap detectors. Muons are detected in gas-ionization chambers embedded in the steel flux-return yoke outside the solenoid. A more detailed description of the CMS detector, together with a definition of the coordinate system used and the relevant kinematic variables, can be found in Ref.~\cite{CMS_ex}.

Events of interest are selected using a two-tiered trigger system~\cite{Khachatryan:2016bia}. The first level (L1), composed of custom hardware processors, uses information from the calorimeters and muon detectors to select events at a rate of around 100\unit{kHz}. The second level, known as the high-level trigger (HLT), consists of a farm of processors running a version of the full event reconstruction software optimized for fast processing that reduces the event rate to around 1\unit{kHz} before data storage.

\section{Data samples and trigger requirements}
\label{sec:samples}

This search uses proton-proton ($\Pp\Pp$) collision data at a center-of-mass energy $\sqrt{s}=13\TeV$ collected by the CMS experiment in the years 2016-2018, corresponding to an integrated luminosity of 138\fbinv. Events that have passed a L1 trigger requirement for the presence of a muon, electron, photon, or jet are selected. The HLT selection requires an event to satisfy at least one of the following criteria:

\begin{itemize} 
\item a muon is present without isolation requirements and with a transverse momentum $\pt > 50 (100)\GeV$ for 2016-2018 (2017-2018) data;
\item an electron is present with isolation requirements on the hadronic energy and with $\pt > 27 (35)\GeV$ for 2016 (2017-2018) data;
\item an electron is present without isolation requirements and with $\pt > 115\GeV$;
\item a photon is present with $\pt > 175 (200)\GeV$ for 2016 (2017-2018) data;
\item the \pt sum of all reconstructed jets is larger than $900 (1050)\GeV$ for 2016 (2017-2018) data.
\end{itemize}

Different trigger thresholds are used because of changes in instantaneous luminosity and detector configuration during the different data-taking periods. 
The inclusion of the photon trigger improves the trigger efficiency for electrons with very large \pt.
The inclusion of the jet triggers improves the efficiency for selecting signal events by approximately 8\% with respect to using lepton triggers alone.
Trigger efficiencies are derived for data and simulation in an event sample using orthogonal triggers requiring two different-flavor leptons, \ie, muon or electron pairs, and enriched in leptonically decaying top quark-antiquark pairs (\ttbar).
Scale factors resulting from the differences between data and simulation are derived as functions of the $\abseta$ and \pt of the lepton and applied as a correction to simulated events.

During the 2016-2017 data taking, a gradual shift in the timing of the inputs of the ECAL L1 trigger in the region $\abseta > 2.0$ caused a specific trigger inefficiency (``ECAL Prefiring"). Dedicated scale factors are applied to simulation in order to account for this effect.

\section{Simulated samples and signal modeling}
\label{sec:samples_sim}

Simulated background samples are generated at LO with \MGvATNLO~\cite{Alwall:2014hca_} for the production of a \PW boson in association with jets (\wjets), and for events with jets arising exclusively from quantum chromodynamics (QCD) interactions (multijet events).
For all these processes, the \MGvATNLO version 2.2.2 (2.4.2) is used for simulated events in conditions corresponding to the 2016 (2017-2018) data taking.
The next-to-LO (NLO) generator \POWHEG 2.0~\cite{Nason:2004rx, Frixione:2007vw, Alioli:2010xd} is used to simulate \ttbar produced in association with jets~\cite{Frixione:2007nw} and single top quark~\cite{fourfiveflavorschemes} events.

The \POWHEG 2.0 and \MGvATNLO generators are interfaced with \PYTHIA  with the following versions and underlying event tunes: version 8.226 and tune CUETP8M1~\cite{Khachatryan:2015pea} for the simulated events used with 2016 data, and version 8.230 and tune CP5~\cite{Sirunyan:2019dfx} for those used with 2017-2018 data. The underlying event tune CUETP8M2T4~\cite{CMS-PAS-TOP-16-021} is employed for the simulated \ttbar events used with 2016 data, and the \PYTHIA version 8.240 is adopted for simulated \ttbar events used for 2017-2018 data. 
The background processes are initially normalized to their theoretical cross sections, using the highest order calculations available.
The cross section for the \ttbar background is computed at next-to-NLO (NNLO) accuracy in perturbative QCD using a soft-gluon resummation at next-to-next-to-leading logarithmic accuracy with the \textsc{Top++} 2.0 program~\cite{Czakon:2011xx}. 
For single top quark production, the NLO cross section as calculated with \textsc{HATHOR} 2.1~\cite{hator} is used. The \wjets cross section is calculated at LO and corrected with an inclusive $k$-factor to account for NLO in electroweak and NNLO in QCD corrections~\cite{Lindert:2017olm}.

Signal event samples are generated at LO using \MGvATNLO version 2.4.2 interfaced to \PYTHIA 8.230. The interactions of the \PWpr boson with the SM particles is described by the Lagrangian term:

\begin{equation}
\mathcal{L}^\text{eff}=\frac{V_{f_if_j}}{2\sqrt{2}}g_\PWpr \bar{f_i} \gamma_\mu [\alpha_{\mathrm{R}}^{f_if_j}(1+\gamma^5) +\alpha_{\mathrm{L}}^{f_if_j}(1-\gamma^5) ]W^{\prime \mu}f_j+\text{h.c.},
\label{eq:lagrangian}
\end{equation}

where $V_{f_if_j}$ is the analogue of the Cabibbo--Kobayashi--Maskawa matrix if $f_i$ and $f_j$ represent quarks, while for leptons  $V_{f_if_j}$ is the identity matrix, and $g_\PWpr$ is the coupling strength of the \PWpr boson. The parameters $\alpha_{\mathrm{R}}^{f_if_j}$ and  $\alpha_{\mathrm{L}}^{f_if_j}$ regulate the coupling strengths to the two chiral fermion components, and depend on the specific gauge model represented in the new physics scenario.

In the following, all $\alpha_{\mathrm{L(R)}}^{f_if_j}$  are varied simultaneously, and we denote them simply as $\alpha_{\mathrm{L(R)}}$.
Samples corresponding to three chirality hypotheses have been generated: RH, with $\alpha_{\mathrm{R}}=1,\alpha_{\mathrm{L}}=0$, LH, with $\alpha_{\mathrm{R}}=0,\alpha_{\mathrm{L}}=1$, or mixed (LR), which consist of equal parts of LH and RH components, specifically $\alpha_{\mathrm{R}}=\alpha_{\mathrm{L}}=1/\sqrt{2}$. The generated \PWpr mass in the samples ranges from 2 to 6\TeV. Narrow-width samples are generated with a width, $\Gamma$, set to 1\% of the resonance mass \mwp in mass steps of 0.2\TeV. Large-width samples are generated with $\Gamma$ set to 10, 20, and 30\% of \mwp, in mass steps of 0.8\TeV. 
All possible final states after top quark decays have been considered for the signal process. 
The signal cross sections are calculated at LO with the \MGvATNLO generator.

Predictions for intermediate values of  $\alpha_{\mathrm{R}}$ and $\alpha_{\mathrm{L}}$ are obtained by a procedure described in Ref.~\cite{CMS:2012nze}.
For an arbitrary combination of $\alpha_{\mathrm{R}}$ and $\alpha_{\mathrm{L}}$, signal distributions, yields, and cross sections are obtained as a linear combination of the ones from LH, RH, and LR \PWpr samples, and the SM single top $s$-channel sample. Values of $\alpha_{\mathrm{R}}$ and $\alpha_{\mathrm{L}}$ between 0 and 1 are considered in steps of 0.1.

{\tolerance=1600
The default parametrization of the PDFs used in all simulations is NNPDF3.0~\cite{Ball:2014uwa} or NNPDF3.1~\cite{NNPDF:2017mvq} at LO or NLO QCD, with the order matching that of the matrix element calculation.
All generated events undergo a full simulation of the detector response according to the model of the CMS detector within {\GEANTfour}~\cite{Agostinelli:2002hh}.
All simulated samples include additional $\Pp\Pp$ interaction in the same or nearby bunch crossings (pileup) and are weighted such that the distribution of the number of interactions in each event agrees with that observed in the data.
\par}

\section{Physics object reconstruction and selection}
\label{sec:event_sel}

The decay chain considered for this analysis is $\PWpr \to \PQt\PQb \to \ell \PGn \PQb\PQb$. Its expected experimental signature consists of one muon or electron, the hadronization products of the two \PQb quarks (\PQb jets), and one neutrino escaping the detector without generating a signal, causing a momentum imbalance in the total \pt of the reconstructed events. 
The experimental signature sought for therefore includes one reconstructed electron or muon, a large amount of missing momentum to the kinematic closure of the events, and at least two jets.
The main backgrounds that mimic this signature are \ttbar and  single top quark processes where a top quark decays through a leptonic decay chain, \wjets production where the \PW boson decays into a lepton and a neutrino, and QCD multijet processes where a fake or nonprompt lepton is reconstructed as a prompt lepton.

After the trigger selection, an offline reconstruction of the events is performed to identify decay vertices and particle candidates.
Particle candidates in the event are reconstructed using the particle-flow (PF) algorithm~\cite{Sirunyan:2017ulk}, which performs a global event reconstruction by combining the information from the various elements of the CMS detector, and which provides the identification of muons, electrons, photons, and charged and neutral hadrons.

The primary vertex is taken to be the vertex corresponding to the hardest scattering in the event, evaluated using tracking information alone, as described in Section 9.4.1 of Ref.~\cite{CMS-TDR-15-02}.
The physics objects considered are the jets formed by clustering the tracks from the candidate vertex by the jet-finding algorithm~\cite{Cacciari:2008gp,Cacciari:2011ma} and the missing \pt (\ptvecmiss) associated with the vertex, taken as the negative vector \pt sum of the jets.
The primary vertex must be within 24\unit{cm} of the nominal interaction point along the beam axis and within 2\unit{cm} in the transverse plane.

Muons considered for this analysis must have $\pt>55\GeV$, and be within the acceptance of the muon system, $\abseta <2.4$.
Electrons considered for this analysis must have $\pt>50\GeV$, be within an $\abseta<2.2$, and pass a multivariate analysis based identification criterion that has an efficiency of 90\% for prompt electrons, \ie, electrons coming from a hard-scattering process, as opposed to nonprompt ones from a hadron decay chain. Electrons in the barrel-endcap gap at $1.444 < \abseta < 1.566$ are excluded from the selection, because the reconstruction of an electron object in this region is not optimal.
In order to distinguish between prompt leptons and the ones coming from hadronic decay chains, isolation criteria are applied based on the deposited energy sum of hadrons and photons in a cone around the lepton direction, compared to the lepton momentum.
For this analysis, an optimized version of the  ``mini-isolation'' variable originally suggested in Ref.~\cite{Rehermann:2010vq} is adopted, tailored for cases where the lepton is produced in a boosted top quark decay. Leptons in signal events often fall within the radius of the jet produced by the \PQb quark arising from the top quark decay, and the separation in the $\eta$--$\phi$ plane between the lepton and the jet axis decreases with the increasing Lorentz boost of the top quark. In order to recover efficiency compared to standard fixed-radius isolation criteria, the variable $\PFminiso{}$ is defined as the ratio between  $S_I(R)$ and $\pt^{\ell}$:
\begin{linenomath*}
\begin{equation}
    \PFminiso{}  = \frac{S_I(R)}{\pt^{\ell}}, \\
    \textrm{with}\,R\,=\frac{10\GeV}{\min(\max(\pt^{\ell },50\GeV), 200\GeV)},
    \label{eqn:riso}
\end{equation}
\end{linenomath*}
where $S_I(R)$ is the scalar \pt sum of charged- and neutral-hadrons, and photon PF candidates inside a cone with a variable radius $R$ around the lepton in the $\eta$--$\phi$ plane and $\pt^{\ell}$ is the \pt of the lepton.
The isolation variable $\PFminiso{}$ is calculated for each lepton, where the cone size decreases with increasing $\pt^{\ell}$ as indicated in Eq.~\eqref{eqn:riso}, reducing the probability of the overlap with a jet for boosted topologies.
For this analysis, muons and electrons are required to have a mini isolation $\PFminiso{}<0.1$.
Events are considered for further analysis if they contain exactly one electron or muon satisfying the requirements listed above. 

In order to reduce the background contribution, particularly from dileptonic \ttbar events, events with at least one additional muon  with $\pt> 35\GeV$,  $\abseta <2.4$, and $\PFminiso{}<0.4$ or at least one additional electron  with $\pt>35\GeV$, $\abseta<2.2$, and $\PFminiso{}<0.4$ are discarded.

Scale factors accounting for the  differences between the lepton identification and mini-isolation efficiencies in data and simulation are derived using a $\PZ \to \ell\ell$ sample, as functions of the $\abseta$ and \pt of the lepton, and applied as corrections to simulated events.

Jets are reconstructed with the anti-\kt algorithm~\cite{Cacciari:2008gp} as implemented in the \FASTJET package~\cite{Cacciari:2011ma}. Two types of jet definitions are used in the analysis:

\begin{itemize}
    \item jets clustered with a radius parameter of $R=0.4$ (AK4 jets) are considered for top quark and \PWpr boson candidate reconstruction;
    \item jets clustered with a radius parameter of $R=0.8$ (AK8 jets) are also used in the analysis, first to perform a loose selection, and then in the event categorization to veto hadronic top quark decays coming from SM backgrounds.
\end{itemize}
The anti-\kt algorithms are both ran onto the same set of particle candidates reconstructed by the PF algorithm, and the two resulting collections of jets are not mutually exclusive. The hadronization products of a single quark can therefore be clustered both in an AK4 and in an AK8 jet in the same event.

For the AK8 jets, the ``modified mass drop tagger'' algorithm~\cite{Butterworth:2008iy,Dasgupta:2013ihk}, also known as the ``soft drop'' (SD) algorithm, with angular exponent $\beta = 0$, soft cutoff threshold $z_{\text{cut}} < 0.1$, and characteristic radius $R_{0} = 0.8$~\cite{Larkoski:2014wba}, is applied to remove soft, wide-angle radiation from the jet.

The charged hadron subtraction algorithm~\cite{Cacciari:2007fd} is applied to the AK4 jets to remove charged hadrons not originating from the primary vertex, while for the AK8 jets the pileup-per-particle identification algorithm~\cite{Bertolini:2014bba} is employed, which assigns a weight to each charged or neutral PF candidate according to the likelihood that the candidate originates from a pileup interaction. 
The weight is then used to rescale the particle four-momentum. 
Correction factors as functions of the \pt, $\eta$, energy density, and the area of the jet are applied to calibrate the jet energy scale. The jet energy resolution for simulated jets is adjusted to reproduce the resolution observed in data~\cite{Khachatryan:2016kdb}. 
Jets potentially coming from instrumental or reconstruction issues are discarded with dedicated selection criteria~\cite{CMS:2017wyc}.

The AK4 jets with $\abseta < 2.4$ and $\pt > 100\GeV$ are included in the analysis, and it is required that at least two of them are present with $\pt > 300$ and 150\GeV for the leading and subleading jets, respectively.
The presence of at least two AK8 jets with $\abseta < 2.4$ and $\pt > 170\GeV$ is also required to reduce contamination from low-energy \ttbar and QCD events, and at the same time allows for the application of a further selection requirement donwstreams in the analysis, described in detail in Section~\ref{sec:categorization}.

A deep neural network based tagger, {\DeepJet}~\cite{Sirunyan:2017ezt,CMS-DP-2018-058,Bols:2020bkb}, is used to identify AK4 jets stemming from the hadronization of \PQb quarks, utilizing information from the tracks, neutral particles, and the secondary vertices within the jet.
The thresholds used for the \DeepJet \PQb tagger in this analysis correspond to a mistag probability for jets initiated by light quarks or gluons with $\pt>500\GeV$ of approximately 5\% in 2016 and 1\% in 2017-2018. The \PQb tagging performance is better in 2017-2018 than in 2016 because of the upgrade of the pixel detector of the CMS tracker in 2017.
This choice of thresholds provides an efficiency of approximately 75\% (60\%)  at $\pt=500\GeV$ and 65\% (50\%) at $\pt=1000\GeV$ for jets initiated by \PQb quarks in the HCAL barrel (endcap) region.
To match the shape of the \DeepJet discriminator in data and simulation, corrections as functions of the \pt and $\eta$ of AK4 jets, derived using samples enriched in dileptonic \ttbar events for jets initiated by \PQb and \PQc quarks, and $\PZ+$jets events for the jets initiated by light quarks and gluons, are applied in simulation. 

The PF-based \ptvecmiss, is the negative vectorial \pt sum of the identified PF particles, and its magnitude is referred to as \ptmiss.
The signal processes are expected to have a significant amount of \ptmiss. A requirement that an event has $\ptmiss > 120\GeV$ reduces the contamination from QCD background processes, which mainly comes from $\PQb\PAQb$ and $\PQc\PAQc$ production, where the \ptmiss originates from nonprompt neutrinos and resolution effects and is thus typically of smaller magnitude.

\section[Top quark and W' boson reconstruction]{Top quark and \PWpr boson reconstruction}
\label{sec:reconstruction}

For events passing the selection defined in Section~\ref{sec:event_sel}, a kinematic reconstruction of top quark candidates from the \PWpr boson decay is performed. The top quark four-momentum is reconstructed starting from the selected lepton, the \ptvecmiss, and one of the AK4 jets considered in the event. 

The top quark momentum is reconstructed according to the algorithm described in Ref.~\cite{Sirunyan:2018rlu}. The reconstruction consists of the following steps:
\begin{itemize}
\item the \ptvecmiss is taken as the transverse component of the neutrino momentum, $\ptvec^\PGn$;
\item the longitudinal component of the neutrino momentum, $\pznu$, is calculated imposing the mass of the lepton-neutrino pair is equal to the \PW boson mass~\cite{PDG2022}. The masses of the lepton and the neutrino are taken to be zero, resulting in a second order equation for $\pznu$.  An additional step is needed if the equation admits two real solutions or two imaginary solutions;
\item if two real solutions are found, the one that results in a value of the \PW boson mass closest to the world average value is chosen. Since the reconstructed leptons have a very small, but nonzero mass, the resulting reconstructed \PW boson mass will not be equal to the imposed SM value;
\item if two imaginary solutions are found, the constraint $\ptvec^\PGn=\ptvecmiss$ is released. This allows the discriminant of the second-order equation to be set to zero, finding a single solution for $\pznu$, and one constraint on the transverse components of the neutrino. To fully determine the neutrino \pt, an additional condition is set by minimizing the vectorial distance between $\ptvec^\PGn$ and $\ptvecmiss$.
\end{itemize}

Given that there are at least two AK4 jets in each event, multiple top quark candidates can be constructed. It is therefore necessary to define a criterion to choose the jet assigned to the top quark, henceforth referred to as \jettop.

Taking into account the kinematic features of the decay chain, the following three criteria are considered in performing the jet assignment:

\begin{itemize}
\item \textbf{mass criterion}: since the \PW boson and the candidate jet are products of the top quark decay, one expects the invariant mass of the sum of the corresponding four-momenta to be close to the world-average value of the top quark mass~\cite{PDG2022}. For the mass criterion we therefore choose, as \jettop, the one that results in a reconstructed top quark mass (\mtop) closest to the nominal value.
\item  \textbf{closest criterion}: we choose, as \jettop, the jet with the lowest angular separation $\Delta R$ from the lepton in the event.
\item  \textbf{subleading criterion}: one expects the highest \pt AK4 jet (the ``leading jet'' ) to arise from the \PQb quark coming directly from the \PWpr boson decay, while the second jet in \pt (the ``subleading jet'') should originate from the \PQb quark coming from the \PQt quark decay. This criterion therefore chooses the subleading jet for the reconstruction of the \PQt quark.
\end{itemize}

If the same AK4 jet satisfies at least two of these criteria, it is chosen as \jettop and the top quark four momentum is reconstructed. If three different AK4 jets are selected by the three criteria, the one passing the mass criterion is selected. 

In order to reconstruct the \PWpr boson candidate in the event, another AK4 jet is needed,  referred to as \jetwprime. This AK4 jet is chosen as the one with the highest \pt among the jets not chosen as the \jettop.

If there are fewer than two \PQb-tagged jets, the \jettop and \jetwprime selection procedure described above is applied to the entire jet collection. If there are at least two \PQb-tagged jets, the \jettop and \jetwprime selection procedure is applied only to the \PQb jet collection.

The reconstructed \mwp, \mlvjj, with $\ell = \Pgm,\Pe$, is obtained as the invariant mass of the reconstructed top quark and \jetwprime four-momenta. 
The quality of the reconstruction algorithm is evaluated on a simulated signal sample with an \mwp of 4\TeV, and it is assessed that the correct jet is assigned in 91\% of the cases for the top quark, and in 87\% of the cases for the \PWpr boson.

\section{Event categorization}
\label{sec:categorization}
Events are divided into four regions, depending on the number of \PQb tagged jets, and whether they are selected as \jettop or \jetwprime. If an event has at least two \PQb-tagged jets, it belongs to the \rtwob signal-enriched region. Conversely, if there are fewer than two \PQb-tagged jets, the region is labeled as \rz, a control and validation region with no signal contamination, if none of \jettop and \jetwprime are \PQb-tagged jets, \rt, a signal-enriched region, if only the  \jettop is \PQb-tagged, and  \rw, another signal-enriched region, if only the  \jetwprime is \PQb-tagged.  The four regions are described in Table~\ref{tab:categories}.

\begin{table}[htbp!]
\begin{center}
\caption{ The regions defined in the analysis depending on the number of \PQb-tagged jets and of the \jettop and \jetwprime assignment. }\label{tab:categories}
{\begin{tabular}{lccc}
Number of \PQb jets & \jettop is a \PQb-tagged jet & \jetwprime is a \PQb-tagged jet & Type of category (label)\\ \hline
0 & no & no & Control region (\rz)\\
\multicolumn{4}{c}{Signal-enriched regions}\\ \hline
1 & yes & no & \PQt jet region (\rt)\\
1 & no & yes & \PWpr jet region (\rw)\\
$\ge$2 & yes & yes & Region with 2 \PQb jets (\rtwob)\\
\end{tabular}}
\end{center}
\end{table}

After the event is assigned to one of the regions, events are further categorized in so-called subregions. Two variables are used in this step:
\begin{itemize} 
\item  the soft-drop mass \msd of the AK8 jet (\jetAKeight) with the smallest angular distance in the $\eta$--$\phi$ plane, $\Delta R(\jetwprime,\jetAKeight)$, to the \jetwprime. If $\Delta R(\jetwprime,\jetAKeight)<0.4$, the soft-drop mass of this \jetAKeight is taken to be \msd. If $\Delta R(\jetwprime,\jetAKeight)>0.4$, the event is not retained for further analysis.
\item the reconstructed top quark mass \mtop, defined as the invariant mass of the reconstructed top quark four-momentum vector, obtained with the procedure described in Section~\ref{sec:reconstruction}.
\end{itemize}

For each of the regions of Table~\ref{tab:categories}, four subregions are identified and labeled using subscripts $\mathrm{A}$, $\mathrm{B}$, $\mathrm{C}$, and $\mathrm{D}$, as defined below, and illustrated in Fig.~\ref{fig:ControlRegions}:

\begin{itemize}
\item subregion $\mathrm{A}$ requires $\msd < 60\GeV$ and $120 < \mtop < 220\GeV$; 
\item subregion $\mathrm{B}$ requires $\msd < 60\GeV$, $\mtop > 220\GeV$,  in \rz, \rw, and \rt, and $\msd < 30\GeV$, $\mtop > 340\GeV$  for \rtwob. This different selection in \rtwob prevents signal contamination; 
\item subregion $\mathrm{C}$ requires $\msd > 60\GeV$ and $120 < \mtop < 220\GeV$;
\item subregion $\mathrm{D}$ requires $\msd > 60\GeV$ and $\mtop < 120$ or $\mtop > 220\GeV$. 
\end{itemize}

The subregions \rtwoba, \rta, and \rwa are used for signal extraction. In such subregions, the requirement on $\msd$ reduces contamination from \ttbar events decaying into a $\ell+\text{jets}$ final state, where the \jettop likely comes from a genuine leptonically-decaying top quark, and the \jetwprime is likely to come from a hadronically decaying top quark. For hadronically decaying, highly energetic top quarks, the decay products are often reconstructed as a single AK8 jet whose soft-drop mass is close to the top quark mass. For the same subregions, the requirement on \mtop reduces contamination from \wjets and multijet QCD events, where a genuine leptonically-decaying top quark is not present in the event, as well as from top quark candidates from \ttbar events in the $\ell+\text{jets}$ channel where the \PQb jet association failed, or from residual dilepton events that happened to pass the lepton veto.

The subregions \rza and \rwa are dominated by \wjets events, while \rta and \rtwoba are dominated by \ttbar events, with some contributions from \wjets, single top quark, and multijet QCD events.

The subregion \rza is not suitable for signal extraction because of the abundance of \wjets events, with jets coming from light quarks or gluons. For this reason, this subregion is used instead for validation of the background extraction procedure defined below.

Regions B are utilized to determine the overall shape and normalization of the different backgrounds, and Regions C and D to assess, and constraint, systematic uncertainties.

\section{Background estimation}
\label{sec:bkgmodelling}

\begin{figure}[htb]
\centering
\includegraphics[width=.64\textwidth]{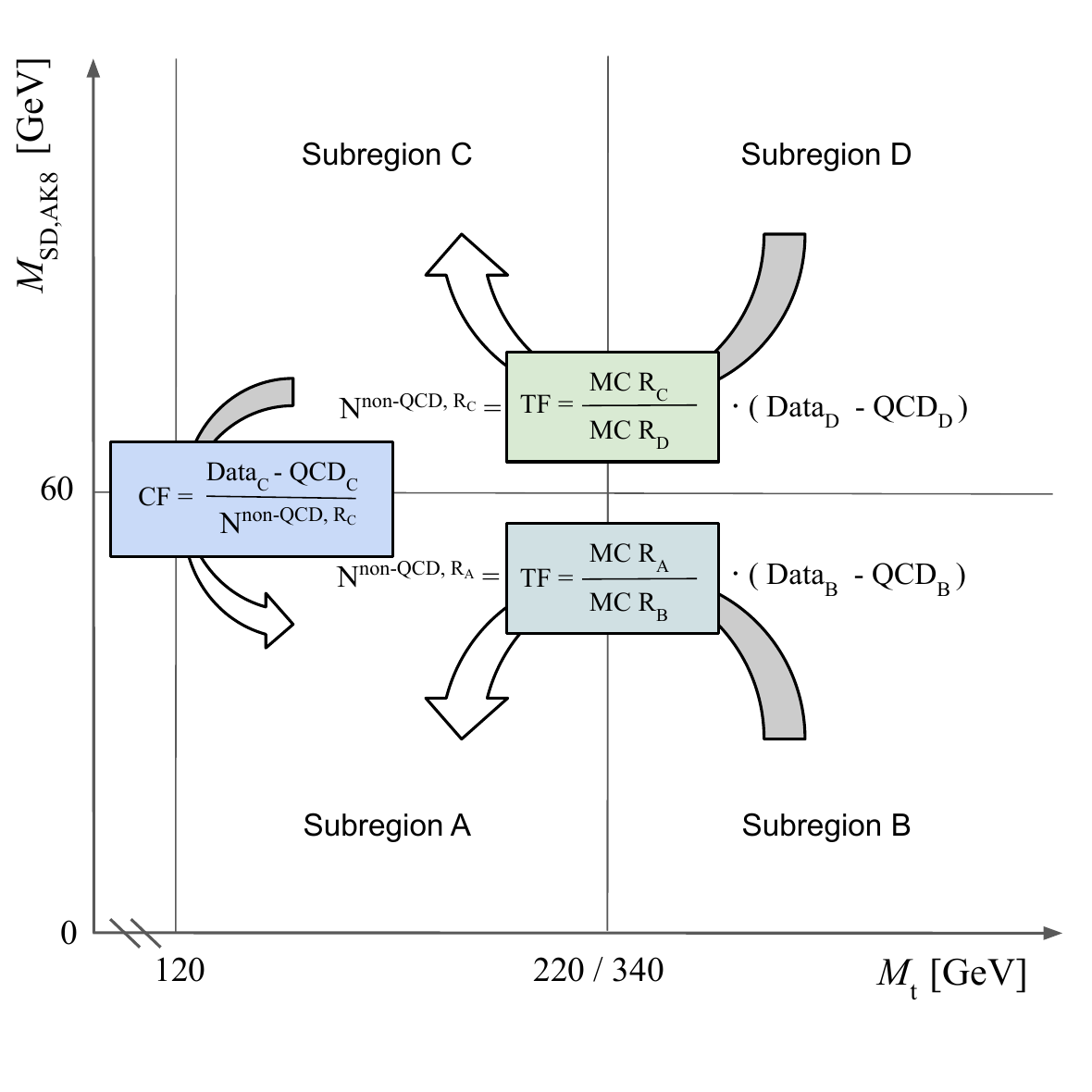}
\caption{A visual representation of the subregions and their usage in the background extraction procedure. The $x$ axis reports the requirements applied on \mtop in order to define the subregions while the $y$ axis represents the soft-drop mass of the AK8 jet associated to the AK4 jet used to reconstruct the \PWpr boson (\msd).}
\label{fig:ControlRegions}
\end{figure}

The modeling of \ttbar, \wjets, and single top quark production, especially at very high values of \mlvjj, is sensitive to higher-order QCD and electroweak corrections that are not included in simulations. In order to correct for these effects, the overall distribution of the background is extracted from the data. The strategy followed in this paper consists in deriving from the entire 2016-2018 data set a single distribution for all the three major background contributions, \ttbar, \wjets, and single top quark production. The multijet QCD process is instead a minor background and simulation is used for its prediction. The background contamination procedure is described in detail in this section while a schematic view is shown in~Fig.~\ref{fig:ControlRegions}.

For each of the three subregions enriched with signal, \rtwoba, \rta, and \rwa, as well as the subregion \rza, all indicated henceforth as ``fit subregions", the corresponding subregion $\mathrm{B}$ is used in order to determine the overall shape and normalization of the reconstructed \mlvjj distribution for the background processes. Simulations are used to determine a transfer function, \TF, from each of those subregions to the corresponding signal-enriched subregion. In the following we describe the procedure in detail, denoting as \ra, \rb, \rc, and \rdd the fit subregion $\mathrm{A}$ and the corresponding subregions $\mathrm{B}$, $\mathrm{C}$, and $\mathrm{D}$, respectively, where \rbig can be either \rtwob, \rt, \rw, or \rz.

The \mlvjj distribution for the sum of the non-QCD simulated background components is considered in \ra and in \rb. The ratio of the two distributions is then modeled with a transfer function that takes the form of:

\begin{equation}
\label{eq:TFModel}
\TF = a\re^{b \mlvjj} + c \mlvjj + d,
\end{equation}

with  $a,b,c,d$ being parameters extracted from a maximum likelihood fit to the Monte Carlo (MC) simulation for non-QCD backgrounds. The following steps are then performed for each bin of the \mlvjj distribution. 

A fit to the \mlvjj distribution in data is performed in the subregion \rb with the same functional form as in Eq.~\eqref{eq:TFModel} with parameters extracted from a maximum likelihood fit. The number of events in each bin as estimated by the fit is then used.
The QCD multijet contribution in \rb is estimated from simulation and subtracted from the resulting distribution. The distribution is then multiplied by the binned transfer function \TFbin from Eq.~\eqref{eq:TFModel} resulting in the background prediction for non-QCD events in the subregion \ra. The overall procedure can be summarized as:

\begin{equation}
\label{eq:Prediction}
N^{\text{non-QCD,\,\ra}} = \TFbin \cdot (N^{\text{Data\,\rb}} - N^{\text{QCD,\,MC\,\rb}}),
\end{equation}
where $N^{\text{non-QCD,\,\ra}}$ is the predicted number of events in the signal subregion \ra, $N^{\text{Data\,\rb}}$ is the number of events estimated from the fit to data in \rb, and $N^{\text{QCD,\,MC\,\rb}}$ is the number of QCD multijet events predicted by simulation in \rb.

Subregions \rc and \rdd are used to assess potential systematic influences on the procedure, either due to the kinematic difference of the \mlvjj spectra induced by the selection on \mtop and not accounted for in simulations, or to the different background composition in the two subregions. 
The background extraction is repeated in \rc and \rdd, by modeling the \mlvjj distribution in data in \rdd with a function of the same form as in Eq.~\eqref{eq:TFModel}. A transfer function $\TF_{\mathrm{CD}}$ between \rc and \rdd is derived from simulation and modeled with a function of the same form as in Eq.~\eqref{eq:TFModel}. The number of events in \rc is obtained as:

\begin{equation}
\label{eq:Prediction_II}
N^{\text{non-QCD,\,\rc}} = \TFbin_{\text{CD}} \cdot (N^{\text{Data\,\rdd}} - N^{\text{QCD,\,MC\,\rdd}}).
\end{equation}
A bin-by-bin correction factor \CF is then defined as:

\begin{equation}
\label{eq:CF_II}
\CF = \frac{N^{\text{non-QCD,\,Data\,\rc}}}{N^{\text{non-QCD,\,\rc}}},
\end{equation}

where $N^{\text{non-QCD,\,Data\,\rc}}$ is the number of events in data in \rc, obtained after subtracting the QCD events in \rc, as determined from simulation.

An alternative shape to the prediction in Eq.~\eqref{eq:Prediction} is obtained as follows:

\begin{equation}
\label{eq:PredictionCorr}
N^{\text{non-QCD,\,corr\,\ra}} = \CF \cdot N^{\text{non-QCD,\,\ra}}.
\end{equation}

The full bin-by-bin difference between the prediction obtained in Eq.~\eqref{eq:Prediction} and the one in~Eq.~\eqref{eq:PredictionCorr} is taken as the systematic uncertainty.

Summarizing the procedure, for each subregion the \TF from \rb to \ra is obtained via a fit to simulation, then multiplied by the distribution in data, which is derived with a fit to \rb in order to soothe statistical effects. To this distribution a small bin-by-bin correction factor is applied as a systematic uncertainty, obtained by repeating the same procedure in \rc and \rd, in order to account for potential biases in the method.

\section{Systematic uncertainties}
\label{sec:systematics}

Several sources of systematic uncertainties have been taken into account. All uncertainties are applied to all background processes are propagated through the background estimation procedure.
\begin{itemize}
\item Luminosity: an uncertainty of $1.6\%$ in the integrated luminosity is used~\cite{CMS-LUM-17-003,CMS-PAS-LUM-17-004,CMS-PAS-LUM-18-002}. This uncertainty is treated as affecting only the number of events, and as correlated across processes and muon and electron channels.
\item Parton distribution functions: the uncertainty due to the PDFs is estimated using reweighted distributions derived from all PDF sets of NNPDF3.1 or NNPDF3.0 according to the PDF4LHC recommendations~\cite{Butterworth:2015oua}.
\item Factorization and renormalization scales: uncertainties related to the choice of these two scales are obtained by considering all variations obtained by setting alternative scales to double or half of the nominal value. The maximum over all variations is taken as the uncertainty. These uncertainties are calculated independently for each process.\item Multijet QCD background: an additional rate uncertainty corresponding to 50\% of the nominal cross section is considered as the uncertainty in the number of QCD multijet events.
\item \PQb tagging and mistagging efficiency: scale factors are applied to simulated events, allowing the reproduction of the \PQb-tagging and mistagging efficiencies, as measured in data. The uncertainty from the scale-factor measurement is propagated to obtain the systematic uncertainty.
\item Pileup modeling: systematic uncertainties related to pileup modeling are taken into account by varying the total inelastic $\Pp\Pp$ cross section of 69.2\unit{mb} by $\pm4.6\%$~\cite{Sirunyan:2018nqx}.
\item Trigger efficiency: data-to-simulation scale factors have been measured in a control sample selected with different trigger requirements with respect to the ones applied for this analysis, selecting a statistically independent event sample. The systematic uncertainty due to the trigger efficiency is obtained by shifting the values of the trigger scale factors up and down by their uncertainties.
\item ECAL prefiring corrections: in 2016-2017, a small fraction of ECAL trigger primitives was associated with a wrong bunch crossing.  Events have been corrected for this effect with a per-event weight, and the corresponding uncertainties have been propagated to the signal yield.
\item Jet energy scale and resolution: in simulated events all reconstructed jet four-momenta are simultaneously varied according to the $\eta$- and \pt-dependent uncertainties in the jet energy scale. In order to evaluate the systematic effect due to differences in the jet energy resolution between data and simulation, a smearing is also applied to simulated events by increasing or decreasing the jet resolutions by their uncertainties~\cite{Khachatryan:2016kdb}. These variations in jet four-momenta are also propagated to the \ptvecmiss.
\item Background modeling, statistical uncertainties: as described in Section~\ref{sec:bkgmodelling}, fits are performed to determine the background shape in the \rb and to obtain the \TF. Statistical uncertainties are propagated to the final distributions. The uncertainties in the parameters obtained from the two fits, to the \TF and to the \mlvjj distribution, are propagated to the predicted number of events in each bin. These uncertainties are treated as uncorrelated across all categories and lepton channels.
\item Background modeling, alternative shapes: alternative models with respect to the form described in Eq.~\ref{eq:TFModel} and Sec.~\ref{sec:bkgmodelling} are considered for the function used to model the data in \rb and for the function used to model the \TF. In this specific scenario, the fit for deriving the number of events starts from 1.5\TeV instead of the nominal starting point of 1\TeV. The fit function is applied within this range to determine the number of events. The resulting uncertainty is initially asymmetric, as it considers the difference between the fit distribution and the nominal distribution. To obtain a two-sided (symmetric) variation, the full difference between the fit distribution and the nominal distribution is considered, but with opposite signs.
\item Background modeling, alternative model: an alternative function is considered for the function used to model the data: the data are described with a Landau function without any further change in the procedure. This uncertainty is obtained by considering the full difference with the nominal distribution and symmetrized to get a two-sided variation.
\item Background composition: when obtaining the \TF, the dependence on the background composition of \ttbar, \wjets, and single top quark production is taken into account by varying each of them independently by a relative factor of $\pm 80\%$ of the expected yield. The entire procedure is repeated by varying each background in all subregions, deriving a new \TF between \ra and \rb, and applying the \TF to the \rb distribution from data. Different alternative templates are obtained with this procedure for each background.
This is done in order to reduce the dependence of the fit on events in the low part of the \mlvjj spectrum, and to separate the low-mass regime from the high-mass regime in the background. This uncertainty is designed to cover differences in the relative fraction of backgrounds among regions, statistical fluctuations in simulation, and the process cross section in the very high-energy regime.
\item Background rate: in order to disentangle the rate and shape effects of the background composition uncertainty, the total background yield for each subregion is left free to float in the fit.
\item Background correction factors: an uncertainty in the correction factor is determined, by considering the full difference between the final distribution with and without the correction (cf. Eqs.~\eqref{eq:Prediction} and \eqref{eq:PredictionCorr}). This uncertainty is then symmetrized by considering the full difference, but with the opposite sign.
\item Limited size of simulated samples: the statistical uncertainty due to the limited size of the simulated event samples is evaluated for each bin with the Barlow--Beeston ``lite'' method~\cite{Barlow:1993dm, bb_light}.
\end{itemize}

The background modeling uncertainties are considered as uncorrelated among regions and lepton channels. This choice is motivated by kinematic and background composition differences among regions and lepton channels. Lepton identification and trigger uncertainties are considered as correlated among regions but as uncorrelated among lepton channels. Experimental uncertainties are considered as correlated among regions and lepton channels. Jet energy scale and resolution uncertainties are uncorrelated among years due to differences in detector conditions. 
Factorization and renormalization scale uncertainties are uncorrelated between all processes, while PDF uncertainties are correlated instead.

Table ~\ref{tab:systematics} shows the range of variation of the leading sources of uncertainty on signal and backgrounds.

\begin{table}[!htb]
\topcaption{Leading systematic uncertainties in the signal and background yields. Two representative signal points are considered, at $\mwp$ = 2 and 6 \TeV, both with $\Gamma/\mwp$ = 1$\%$.  The uncertainties are shown by giving the minimum and maximum across all regions. All values are given as percentages, and uncertainties in the backgrounds are the same for all signal masses.}\label{tab:systematics}
\centering
\cmsTable{\begin{tabular}{lcccc}
    & \multicolumn{2}{c}{ \PWpr signal} & \multicolumn{2}{c}{ Background} \\
Uncertainty & \mwp = 2  \TeV & \mwp = 6  \TeV & Non-QCD background & QCD multijet \\\hline
\wjets component & & & 11--23 & \\
\ttbar component & & & 9--21 & \\
Single top quark component & & & 1--6 & \\
Background shape, statistical & & & 1--5 &  \\
Transfer function, statistical & & & 1--10 & \\
Background shape, modelling  & & & 2--6 & \\
Transfer function, modelling & & & 2 & \\
Background correction factors  & & & 1--9 & \\
\PQb tagging and mistagging & 3--11 & 8--23   & 2  & 4--12 \\
Jet energy scale and resolution & 1 & 1--4 & 2--4 & 10--40 \\  
\end{tabular}}
\end{table}

\section{Statistical inference and results}
\label{sec:results}
A simultaneous maximum likelihood fit to data is performed using the distribution of \mlvjj in both the muon and electron channels. All systematic uncertainties described in Section~\ref{sec:systematics} are treated as nuisance parameters.

\begin{figure}[htb]
\centering
\includegraphics[width=.48\textwidth]{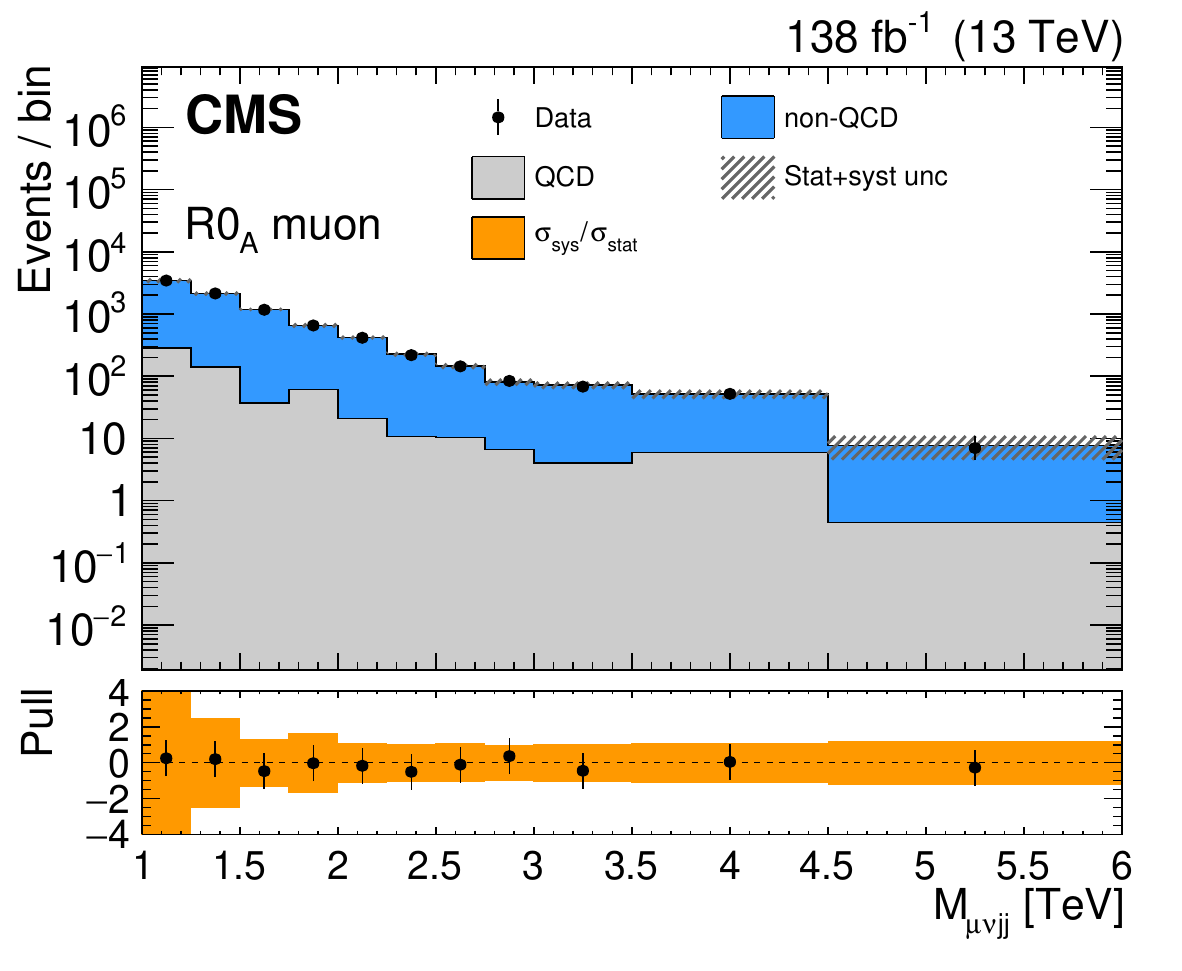}
\includegraphics[width=.48\textwidth]{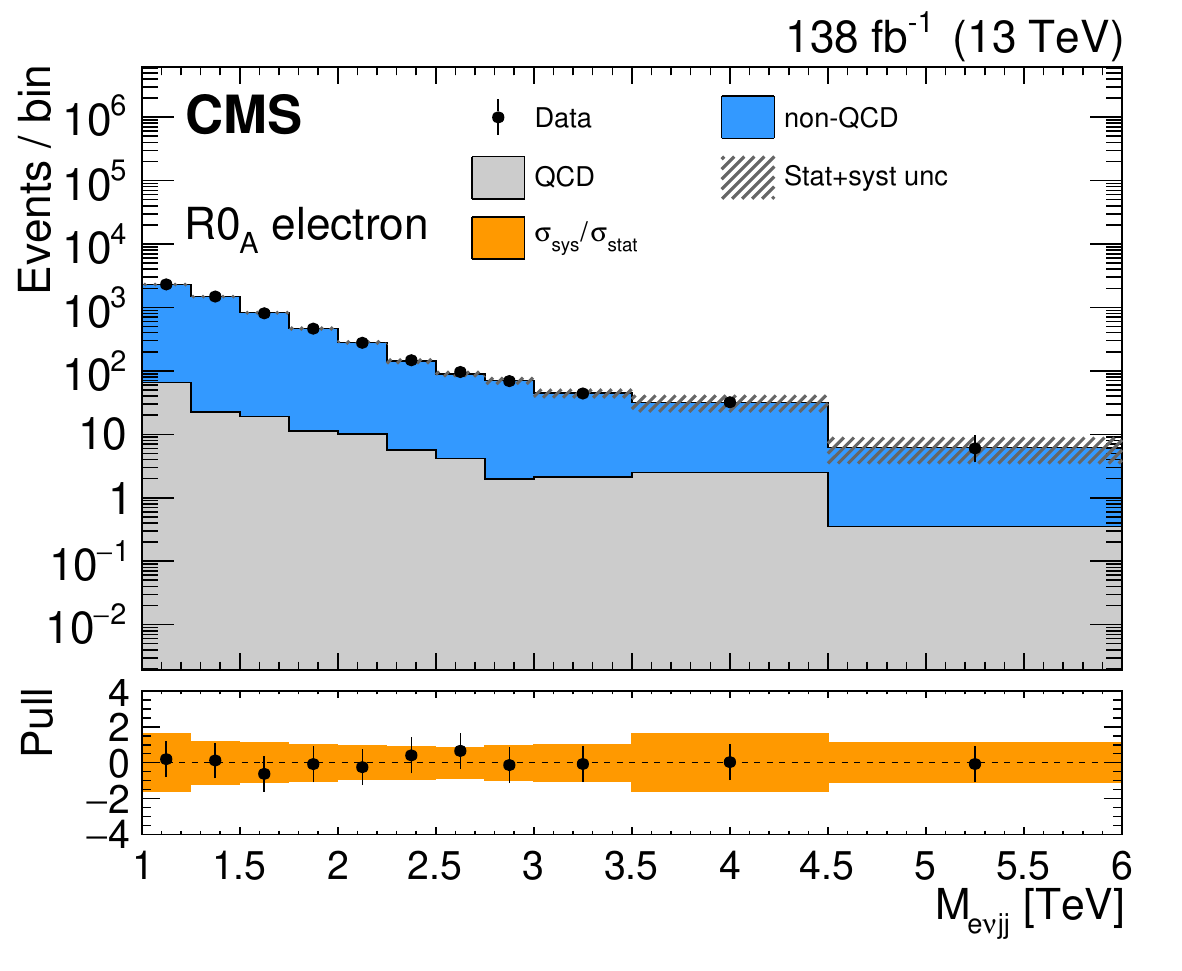}
\caption{Post-fit distributions of \mlvjj in the \rza control subregion for muons (left) or electrons (right). The lower panel reports the data minus the expected number of events normalized to the statistical uncertainty of the data. The orange band represents the systematic uncertainties, also normalized to the statistical uncertainty of the data.}
\label{fig:mwpr_cr0b_postfit}
\end{figure}

A first fit is performed in the subregion \rza in order to validate the background estimation procedure using a statistically independent sample. The post-fit distributions of the reconstructed \mlvjj are shown in Fig.~\ref{fig:mwpr_cr0b_postfit}. 
The distributions are in agreement with the data, as can be seen in the lower panel, where the data minus the expected number of events are presented, normalized to the statistical uncertainty of the data. The orange band represents the systematic uncertainties, also normalized to the statistical uncertainty of the data. In addition, a goodness of fit test was conducted, yielding a $p$-value of 0.128~\cite{Gross:2010qma}. Signal injection tests were performed, ensuring that no significant bias is present due to the statistical inference procedure.

To check for the presence of a possible signal in the data, the fit procedure is carried out simultaneously using the distributions of \mlvjj in the three subregions \rtwoba, \rta, \rwa of the muon and electron channels. 

All signal hypotheses described in Section~\ref{sec:intro} are probed in separate fits, including purely left- and right-handed chiralities of the \PWpr boson, and width hypotheses of 1, 10, 20, and 30$\%$ of the \mwp over a range between 2-6\TeV. 
For left-handed and mixed scenarios, the analysis is performed by considering SM $s$-channel single top quark production as a signal in the statistical inference procedure.

\begin{figure}[htbp!]
\centering
\includegraphics[width=.48\textwidth]{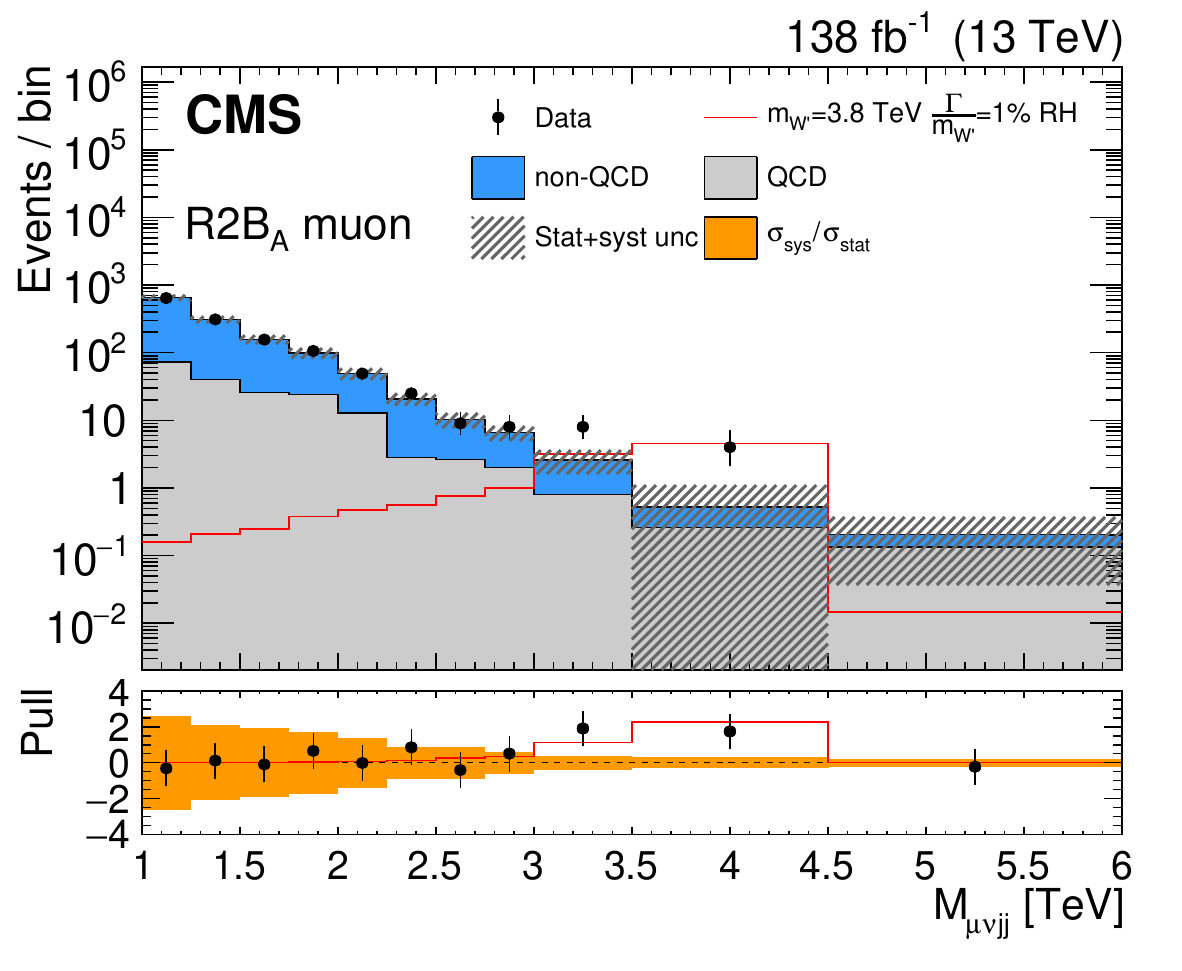}
\includegraphics[width=.48\textwidth]{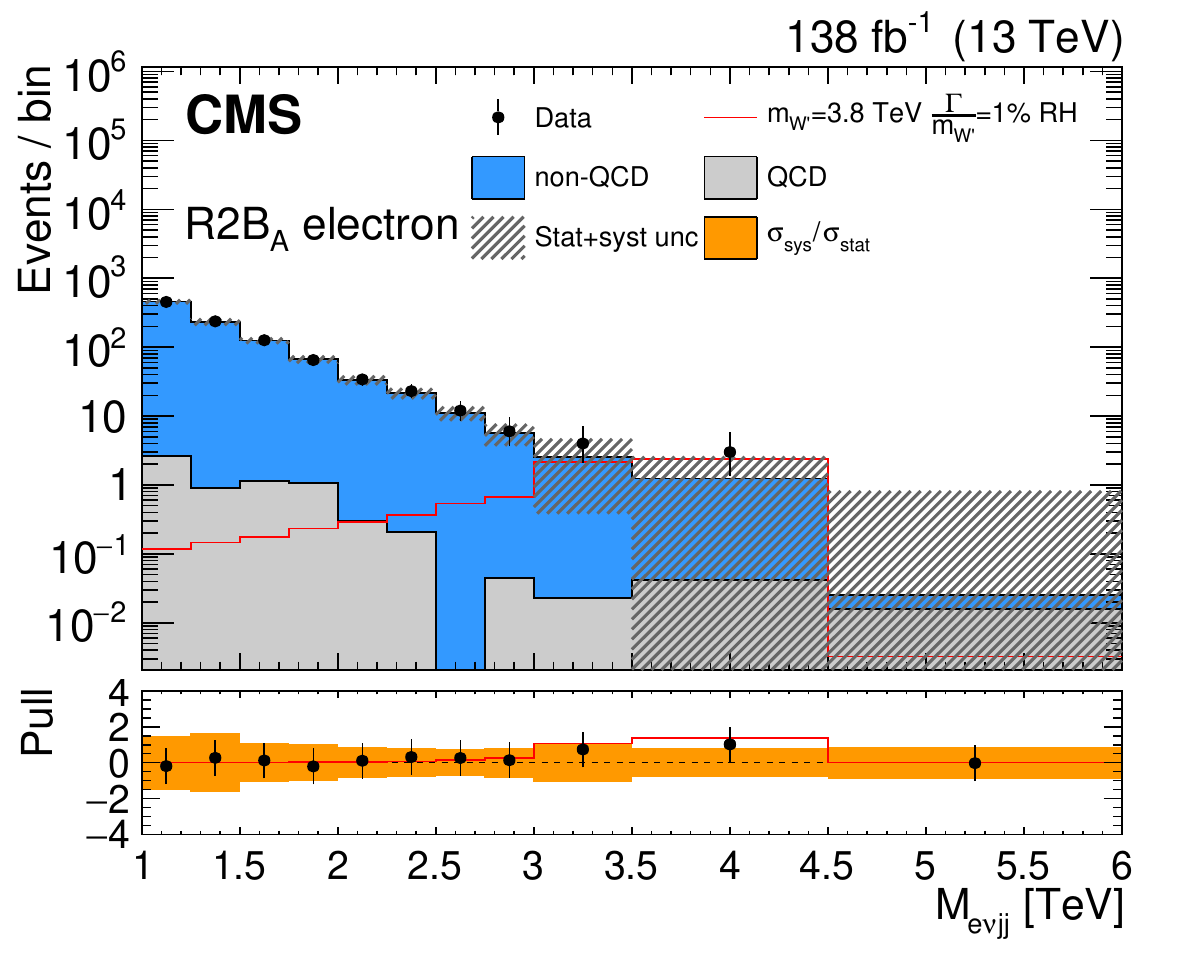}\\
\includegraphics[width=.48\textwidth]{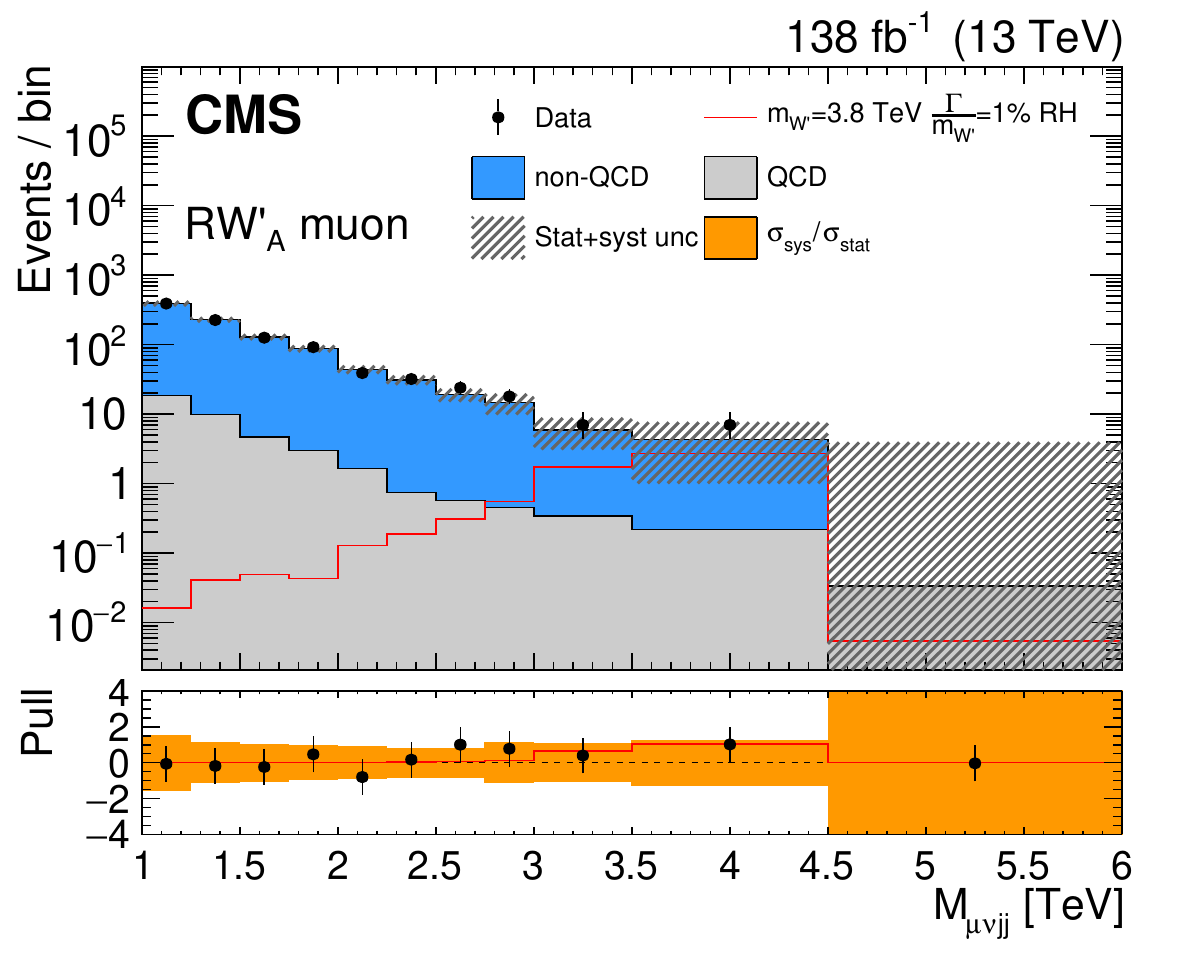}
\includegraphics[width=.48\textwidth]{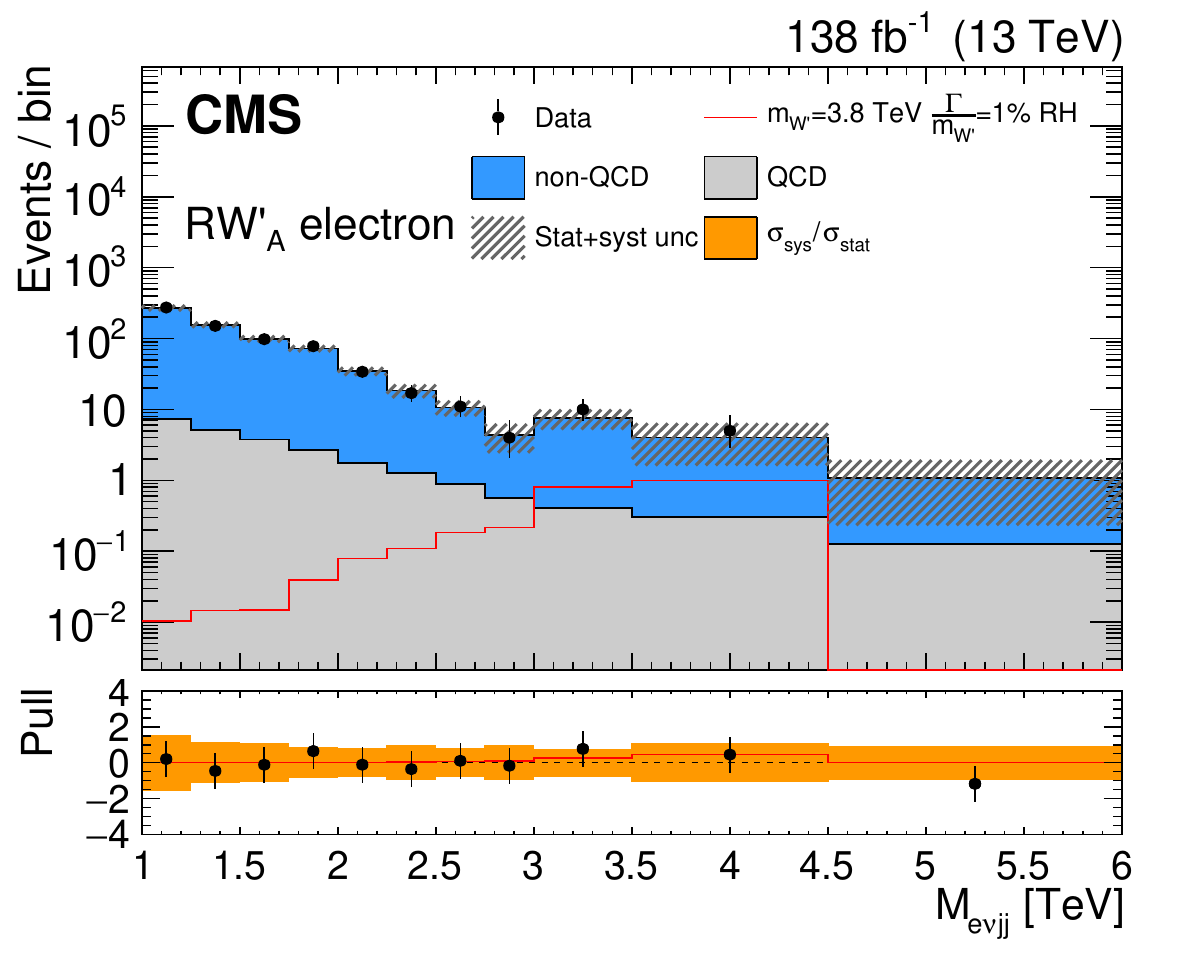}\\
\includegraphics[width=.48\textwidth]{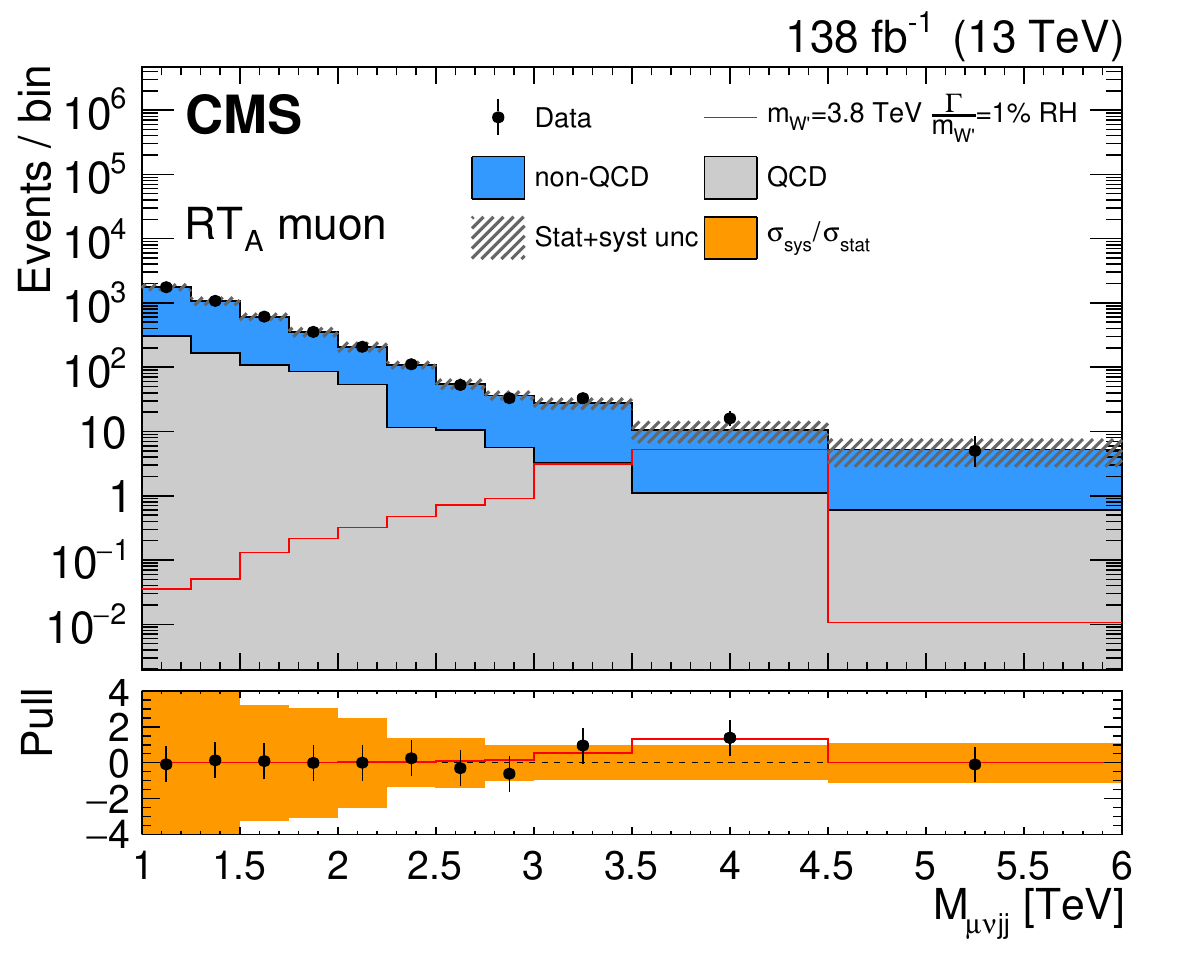}
\includegraphics[width=.48\textwidth]{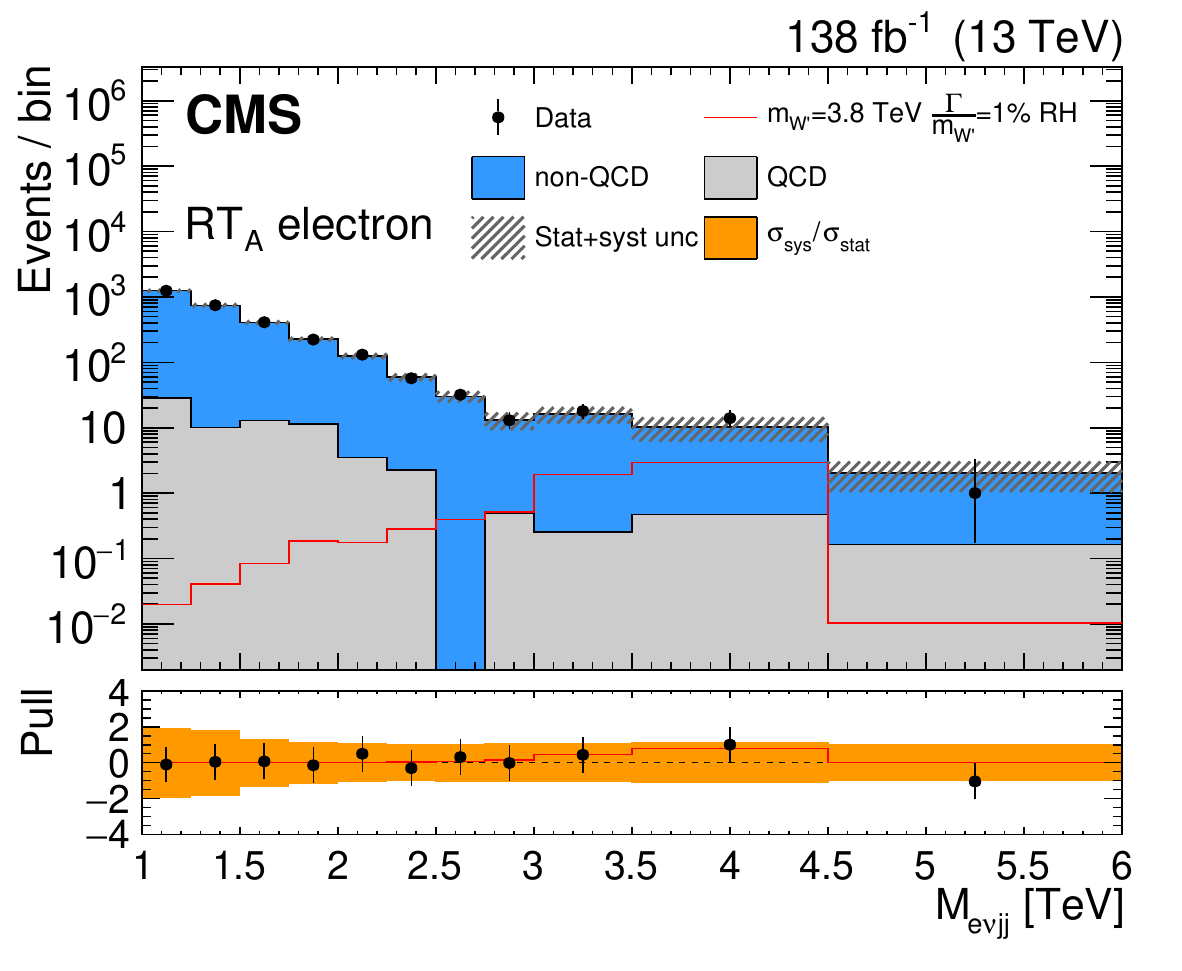}
\caption{Post-fit distributions of \mlvjj in the \rtwoba (upper), \rwa (middle), and \rta (lower) subregions for muons (left column) and electrons (right column). All process yields and nuisance parameters are set to the values obtained from the background plus signal fit. The signal considered for the fit corresponds to the purely right-handed production of a \PWpr with $\mwp$ of 3.6 \TeV and a relative width of 1$\%$ of the $\mwp$, and is represented by the solid red line.
 The lower panels show the data minus the expected number of events, normalized to the statistical uncertainty of the data. The orange band represents the systematic uncertainties, also normalized to the statistical uncertainty of the data.}
\label{fig:mwpr_sr_postfit}
\end{figure}

A local excess over the background is observed for masses $\mwp$ between 3.4-4.4\TeV, with a maximum local significance of 2.6 standard deviations at 3.8\TeV for a 1$\%$ relative width, and a right-handed signal hypothesis. The most significant contribution to the excess comes from the \rtwoba region for the muon channel, making it the most sensitive subregion in the analysis. An important contribution also comes from the \rtwoba in the electron channel and \rta in the muon channel.
The global significance of this excess is 2.0 standard deviations. The corresponding post-fit \mlvjj distributions in the background plus signal hypothesis are shown in Fig.~\ref{fig:mwpr_sr_postfit}. Events in data corresponding to the excess region in \mlvjj were scrutinized, by studying the features of reconstructed physics objects to ensure that all were correctly reconstructed.

The most important systematic uncertainties affecting the result are those related to the \PQb tagging efficiency, the background composition, the background alternative shape, and the background correction factors.

\begin{figure}[htb]
\centering
\includegraphics[width=.48\textwidth]{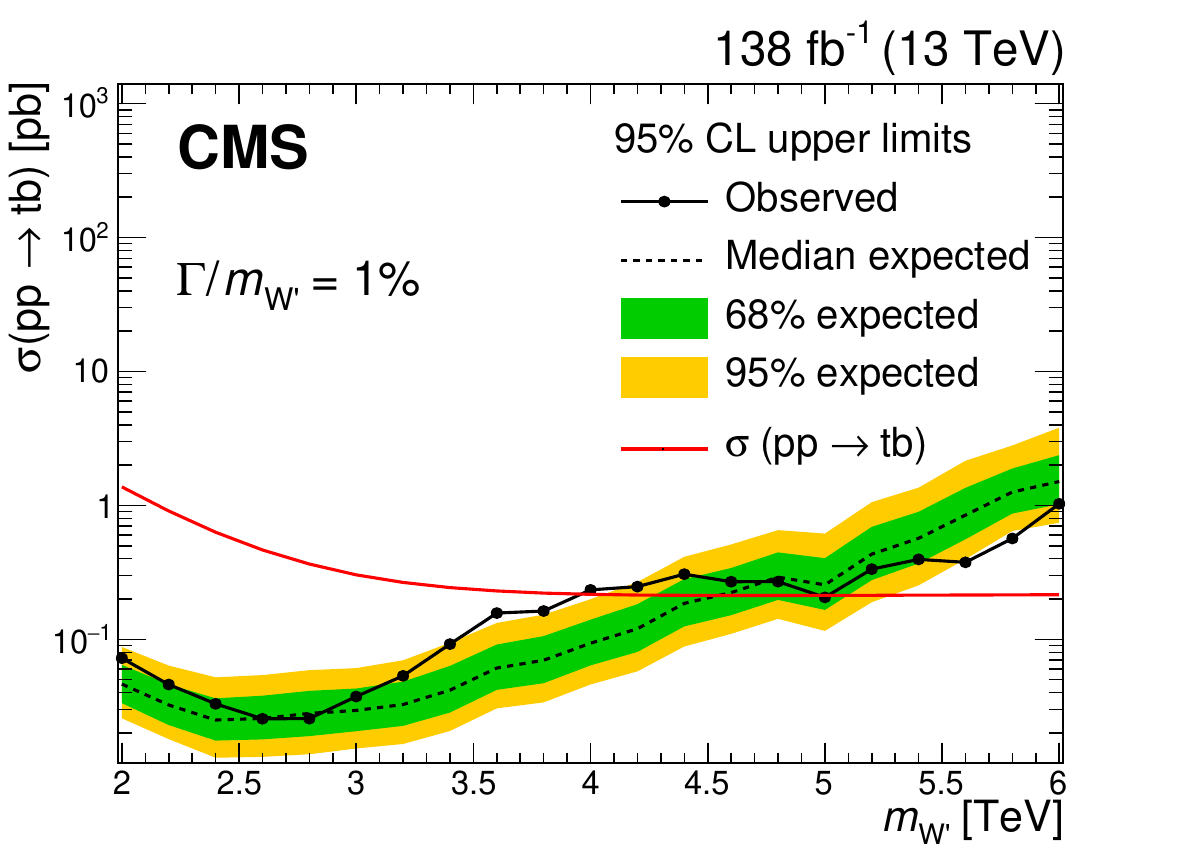}
\includegraphics[width=.48\textwidth]{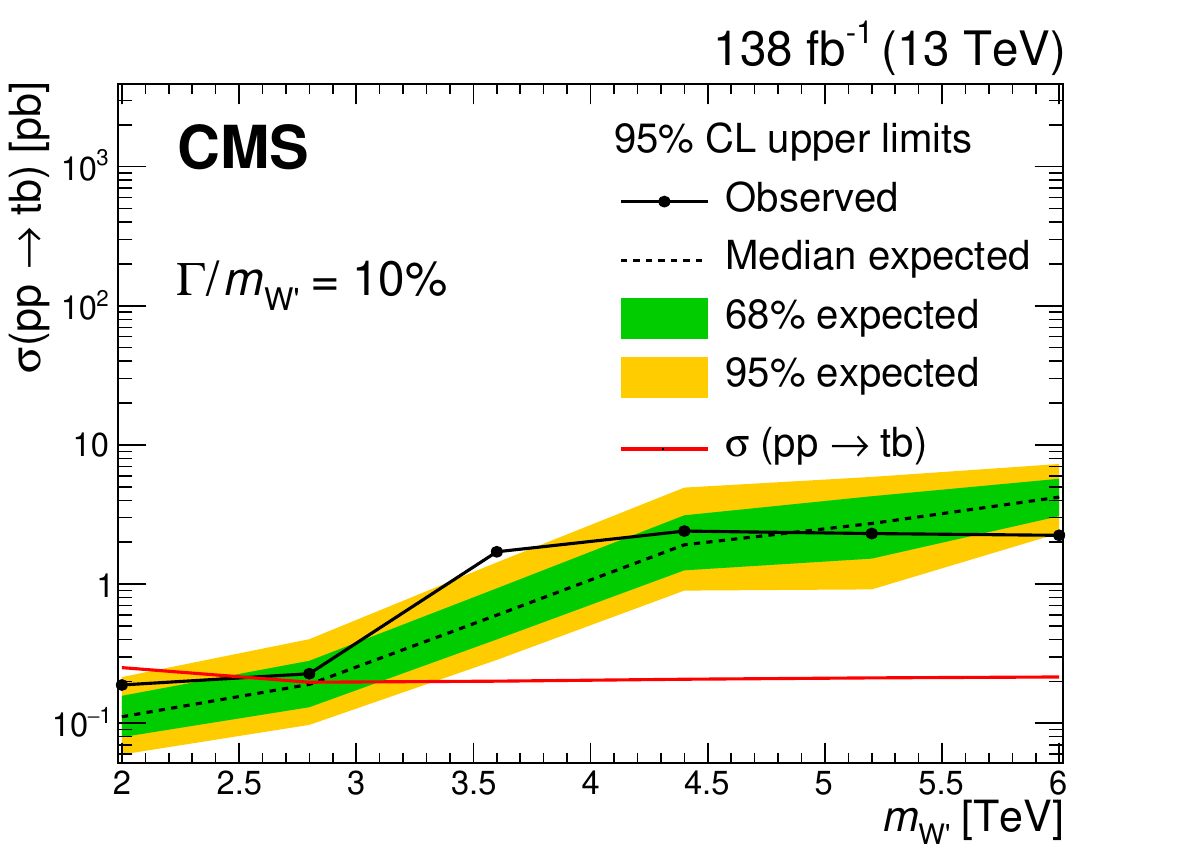}\\
\includegraphics[width=.48\textwidth]{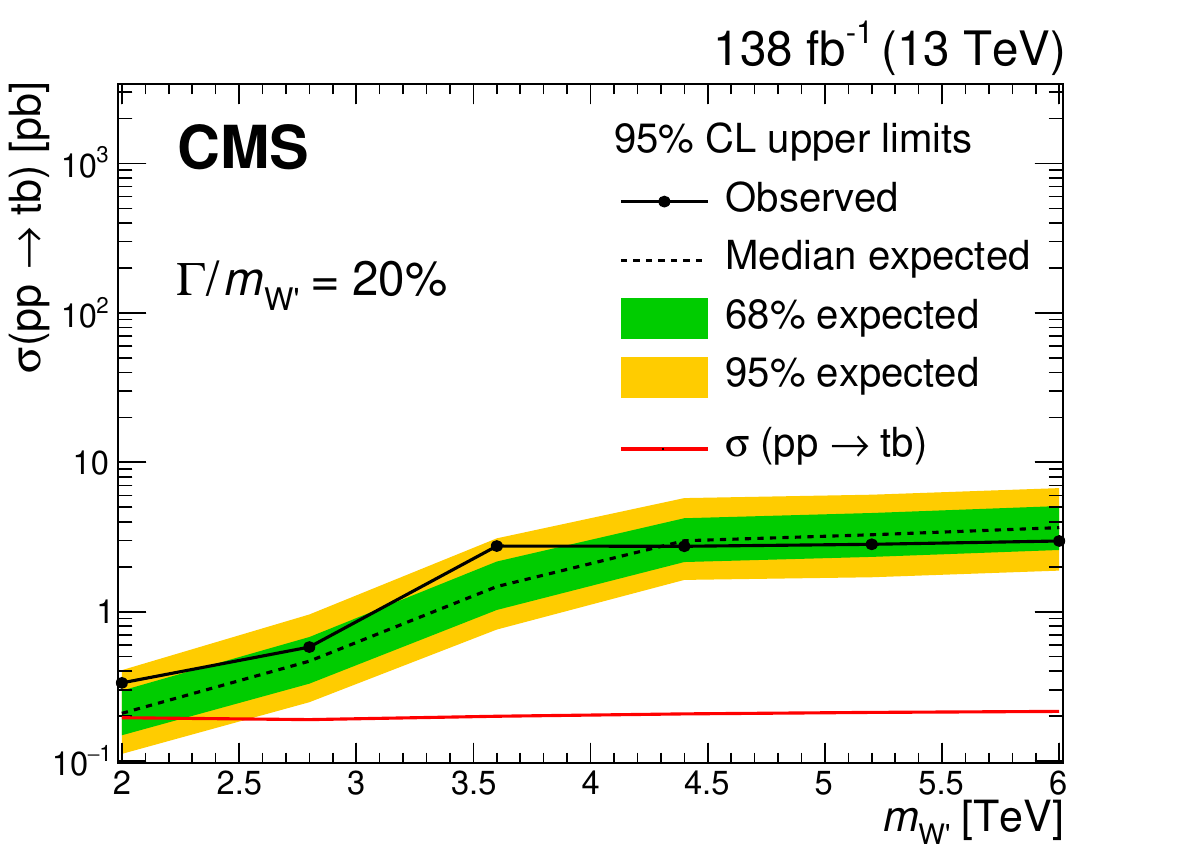}
\includegraphics[width=.48\textwidth]{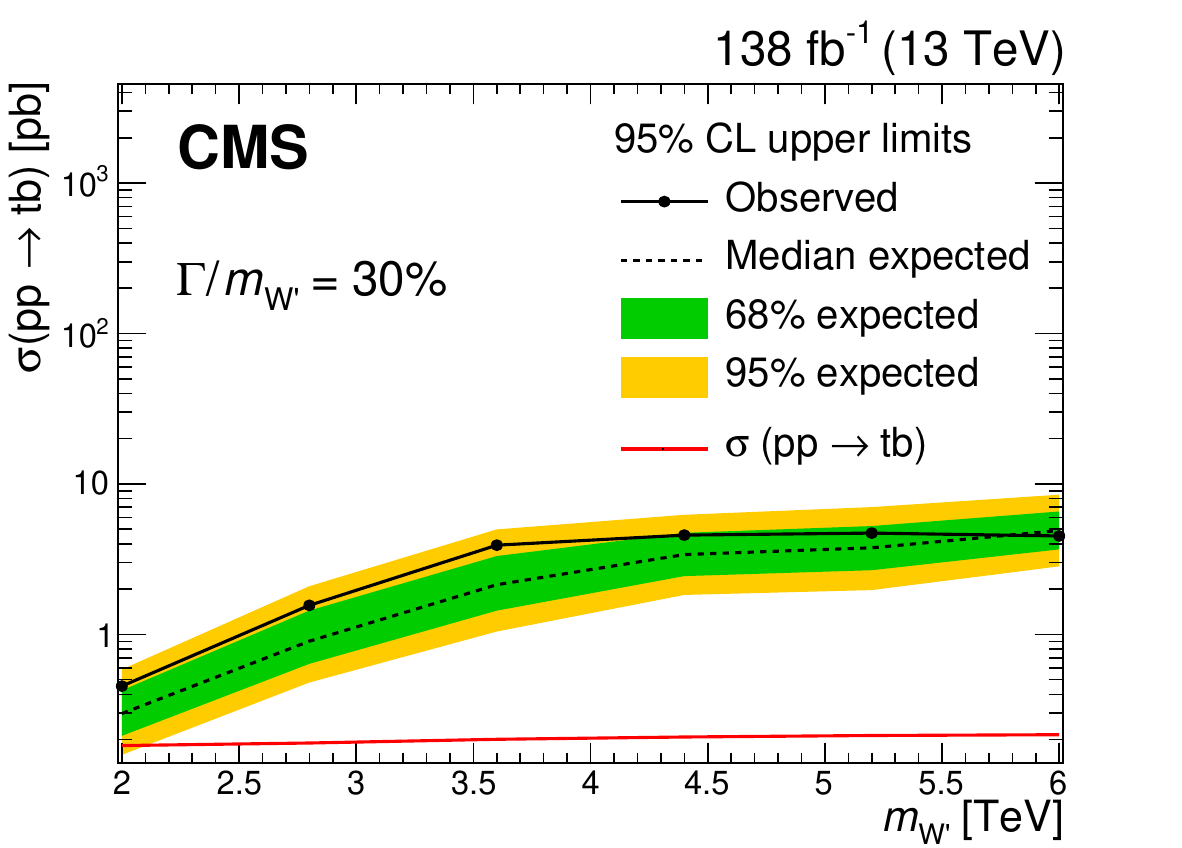}\\
    \caption{Observed and expected 95\% \CL upper limits on the product of the production cross section for production of a $\PQt\PQb$ quark pair in the $s$-channel, mediated by either a \PW or a left-handed \PWpr boson, and including interference terms, given as functions of \mwp for a relative width of 1\% (upper left), 10\% (upper right), 20\% (lower left), and 30\% (lower right). The inner (green) band and the outer (yellow) band indicate the regions containing 68 and 95\%, respectively, of the distribution of limits expected under the background-only hypothesis. The solid red curves show the theoretical expectation at LO.}
\label{fig:sigma_mass_limits_LH}
\end{figure}

\begin{figure}[htb]
\centering
\includegraphics[width=.48\textwidth]{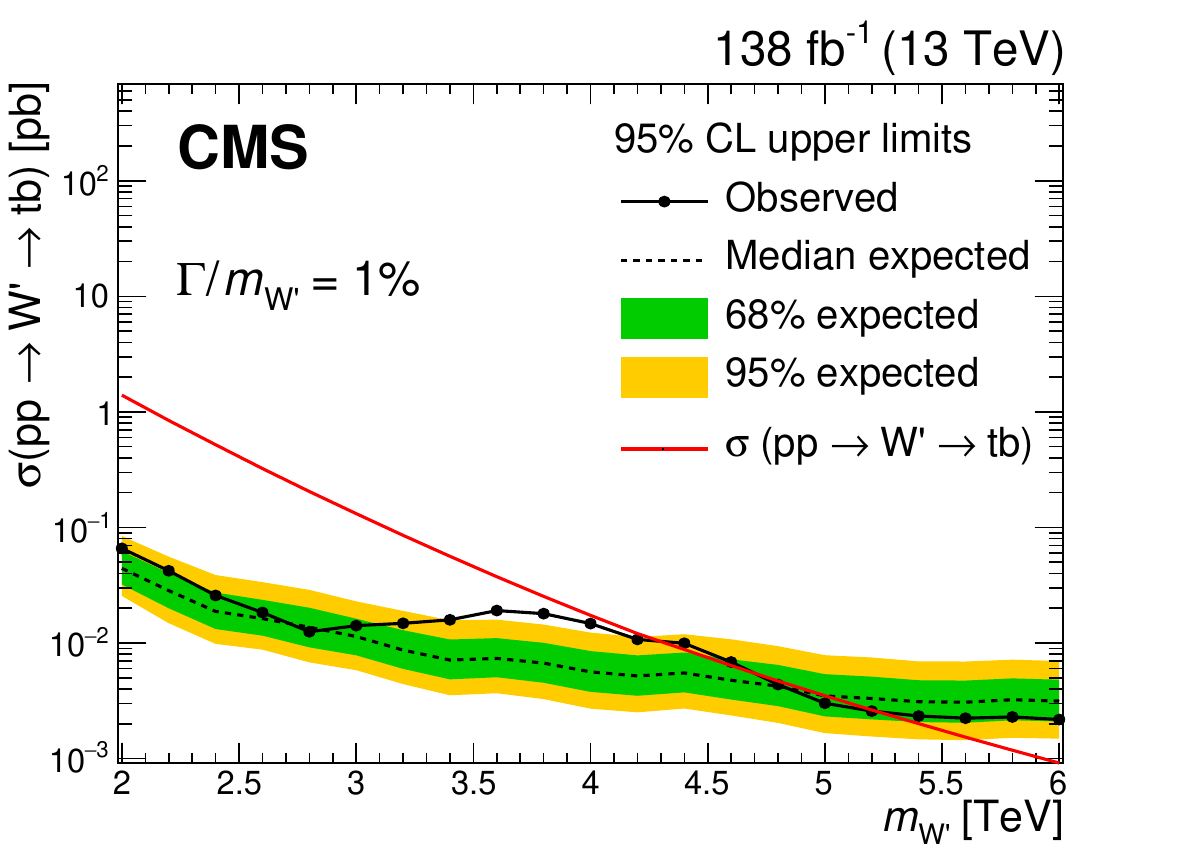}
\includegraphics[width=.48\textwidth]{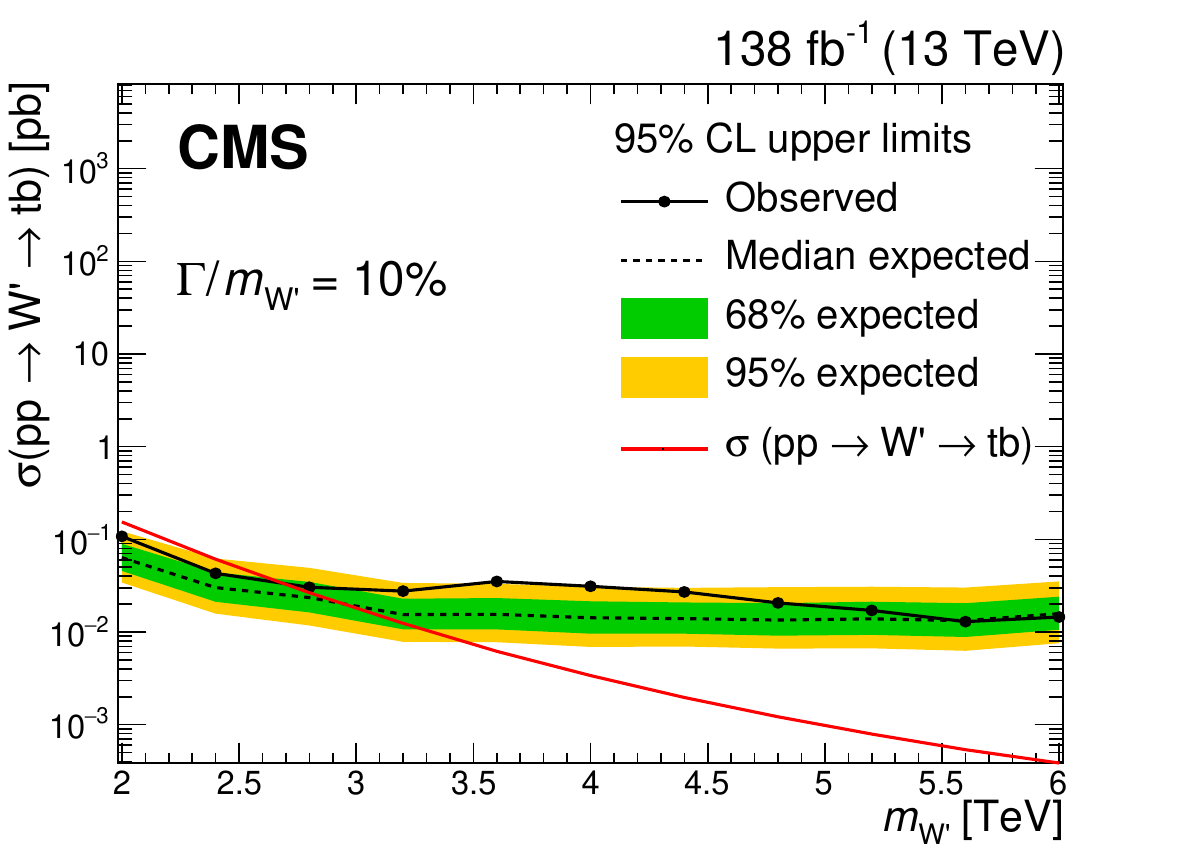}\\
\includegraphics[width=.48\textwidth]{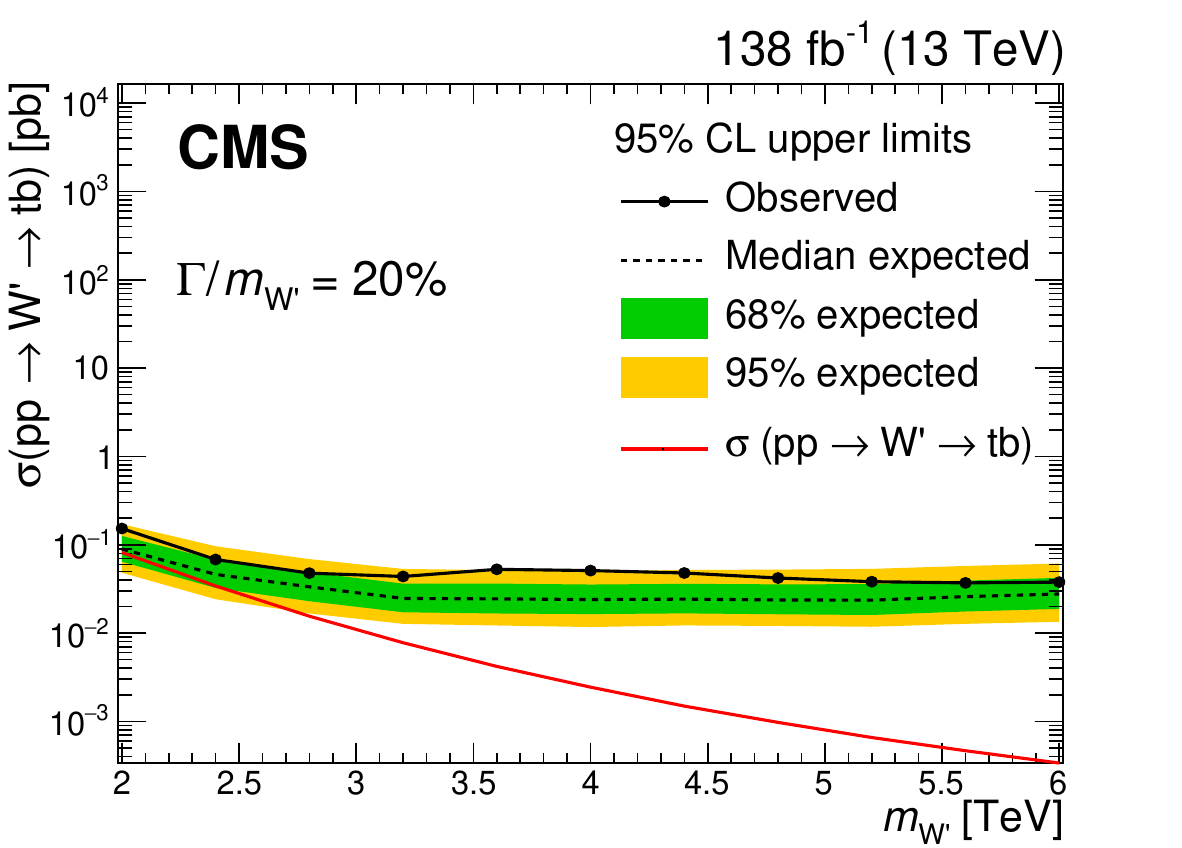}
\includegraphics[width=.48\textwidth]{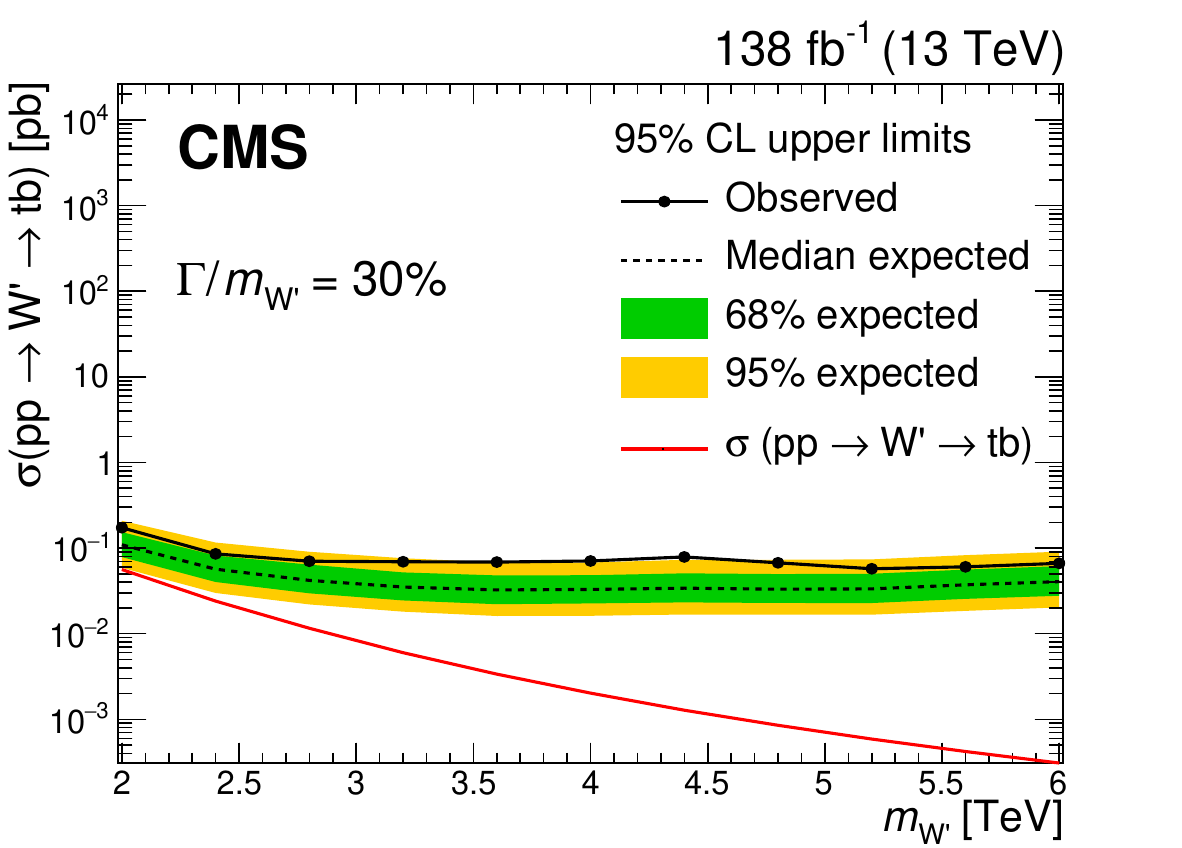}\\
    \caption{Observed and expected 95\% \CL upper limits on the product of the production cross section for a right-handed \PWpr boson and the $\PWpr \to \PQt\PQb$ branching fraction, as functions of \mwp for a relative width of 1\% (upper left), 10\% (upper right), 20\% (lower left), and 30\% (lower right). The inner (green) band and the outer (yellow) band indicate the regions containing 68 and 95\%, respectively, of the distribution of limits expected under the background-only hypothesis. The solid red curves show the theoretical expectation at LO.}
\label{fig:sigma_mass_limits_RH}
\end{figure}

Upper limits on the production cross section are extracted by following the \CLs prescription~\cite{Junk:1999kv,CLs1}, making use of the asymptotic approximation~\cite{Cowan:2010js}. Cross section limits are derived for each value of the \PWpr mass, width, and chirality, and compared with the prediction obtained as described in Section~\ref{sec:bkgmodelling}. 
Figures~\ref{fig:sigma_mass_limits_LH} and~\ref{fig:sigma_mass_limits_RH} show, for LH and RH hypotheses respectively, the upper limits on the cross section as a function of \mwp in the scenarios with $\Gamma/\mwp$ = 1, 10, 20, and 30$\%$. Upper limits for scenarios with $\Gamma/\mwp$ = 10, 20, and 30$\%$ in the case of a fixed branching fraction and a larger coupling are shown in Appendix \ref{sec:alter_xs}.

\begin{figure}[htb]
\centering
\includegraphics[width=.48\textwidth]{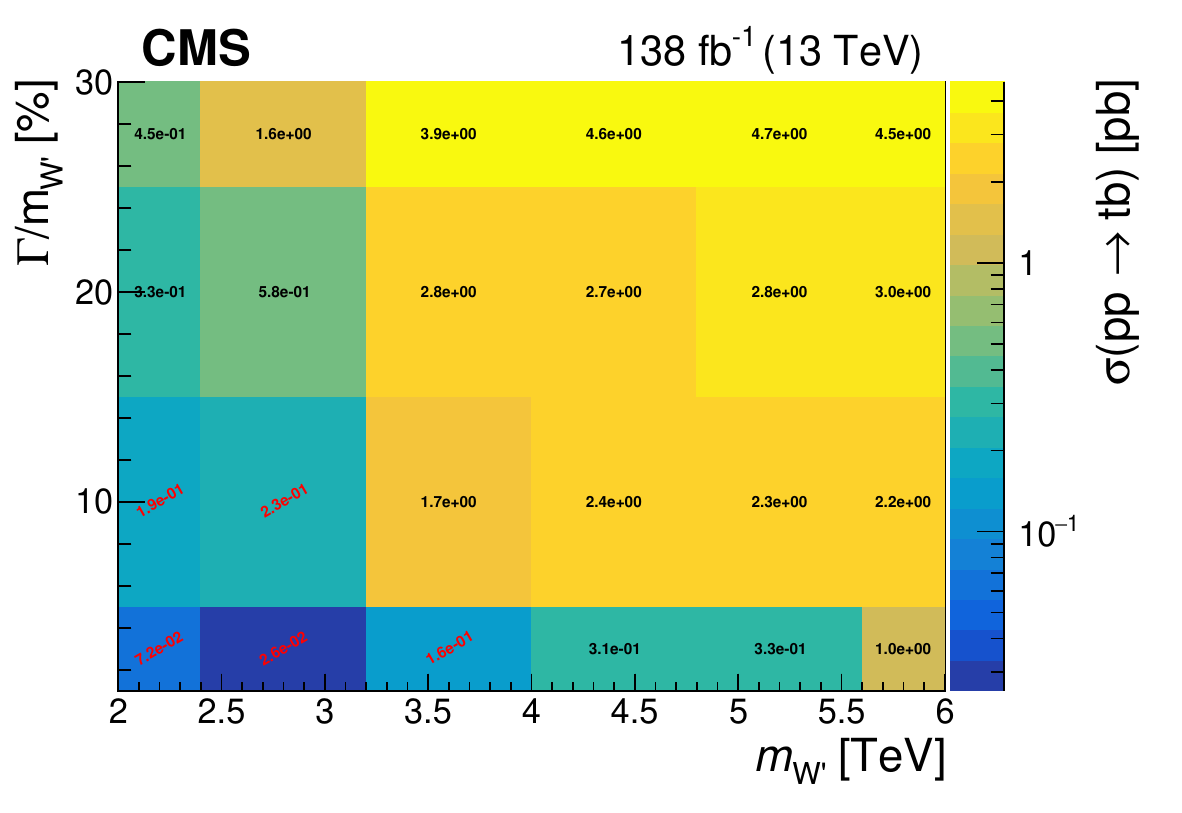}
\includegraphics[width=.48\textwidth]{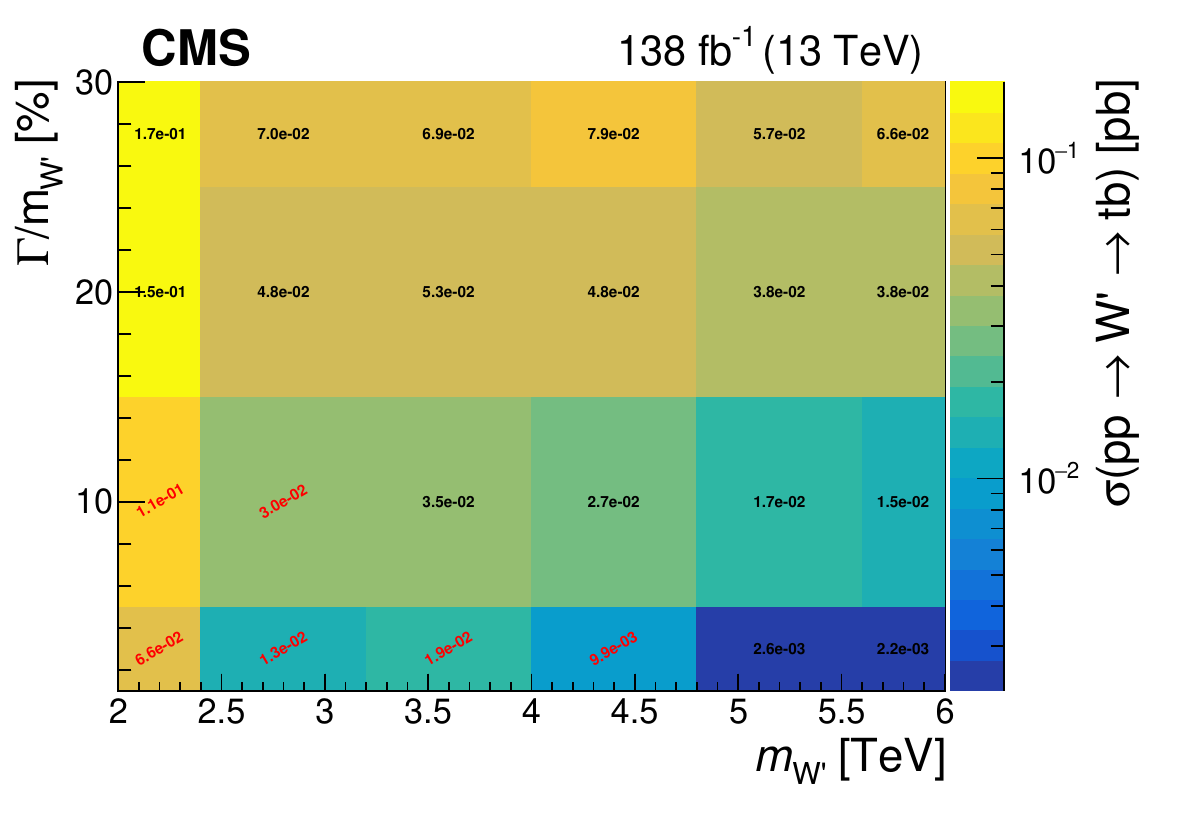}\\
    \caption{Observed 95\% \CL upper limit on the production cross section  for a left- (on the left) and right-handed (on the right) \PWpr boson in the $\PQt\PQb$ final state, as functions of \mwp and relative width \relwidth. Numbers in red, written diagonally, represent values of the excluded cross sections that are lower than the theoretical ones for the analyzed model.}
\label{fig:limit2D}
\end{figure}

The observed 95\% \CL upper limits on the production cross section for a right- and left-handed \PWpr boson in the $\PQt\PQb$ final state are shown in Fig.~\ref{fig:limit2D}, as functions of \mwp and relative width \relwidth. Numbers in red represent values of the excluded cross sections lower than the theoretical ones for the analyzed model.

These limits can be compared with those given in Ref.~\cite{ATLAS:2023ibb}. The analysis presented here provides the most stringent exclusion limits at high masses, for which it was designed, i.e. around 5 TeV and above. At lower masses the limits from Ref.~\cite{ATLAS:2023ibb} are stronger, but they should be compared with the results from Ref.~\cite{CMS:2021mux_}, which are competitive below 4\TeV.

Models with values of the parameters $\alpha_{\mathrm{R}}$ and $\alpha_{\mathrm{L}}$, which regulate the chirality fraction as defined in Eq.~\eqref{eq:lagrangian}, ranging from 0.1 to 0.9 in steps of 0.1 were also tested in the hypothesis of $\Gamma/m_{\PWpr}= 1\%$.
Under the hypothesis of a narrow width, a simple formula can be used to obtain the cross section of the process from the ones for single top $s$-channel production ($\sigma_{\mathrm{SM}}$), pure LH ($\sigma_{\mathrm{L}}$), pure RH ($\sigma_{\mathrm{R}}$), and the mixed LH and RH ($\sigma_{\mathrm{LR}}$) case with $\alpha_{\mathrm{R}}=\alpha_{\mathrm{L}}$~\cite{Boos:2006xe}:

\begin{equation}
    \sigma= (1-\alpha^2_{\mathrm{L}})\sigma_{\mathrm{SM}}+\frac{1}{\alpha^2_{\mathrm{L}} +\alpha^2_{\mathrm{R}}} \left[  \alpha^2_{\mathrm{L}}(\alpha^2_{\mathrm{L}} - \alpha^2_{\mathrm{R}})\sigma_{\mathrm{L}}+\alpha^2_{\mathrm{R}}(\alpha^2_{\mathrm{R}} - \alpha^2_{\mathrm{L}})\sigma_{\mathrm{R}}  +  4\alpha^2_{\mathrm{L}} \alpha^2_{\mathrm{R}}\sigma_{\mathrm{LR}} - 2\alpha^2_{\mathrm{L}} \alpha^2_{\mathrm{R}}\sigma_{\mathrm{SM}} \right].
\end{equation}

\begin{figure}[htb]
\centering
\includegraphics[width=.48\textwidth]{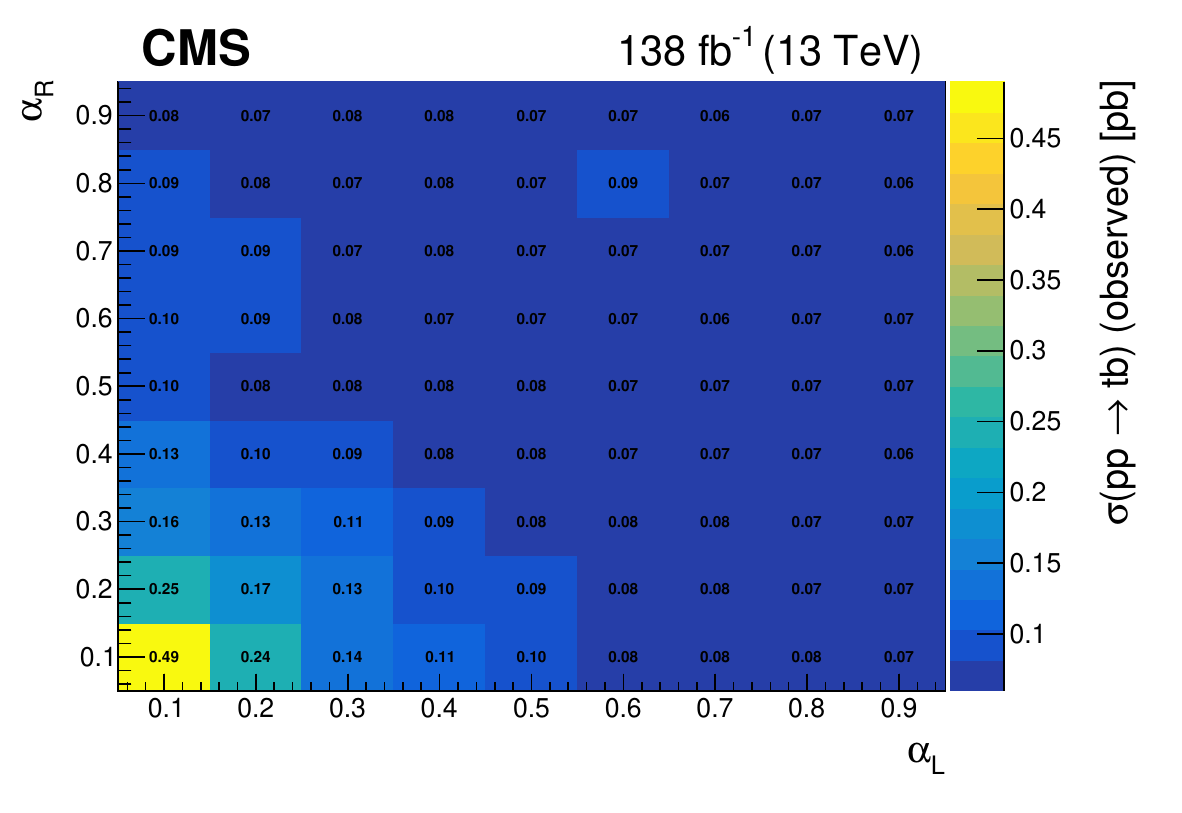}
\includegraphics[width=.48\textwidth]{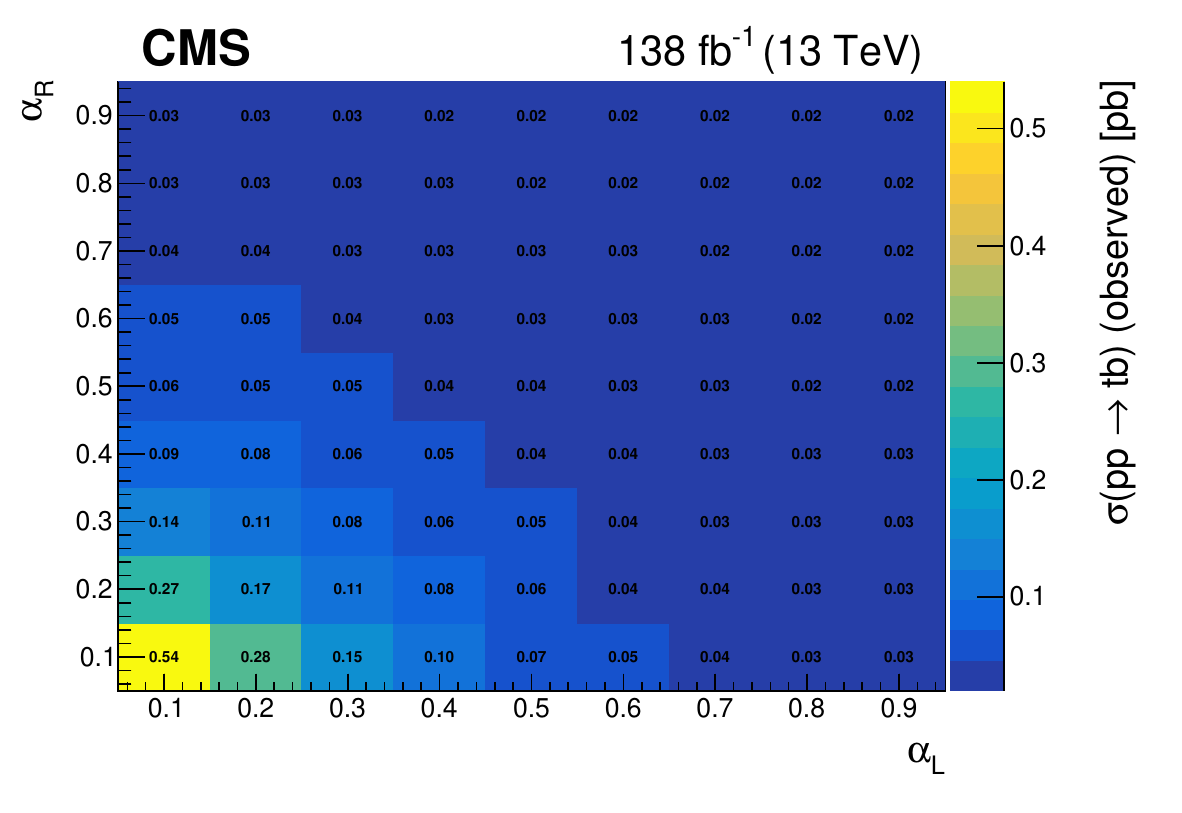}\\
\includegraphics[width=.48\textwidth]{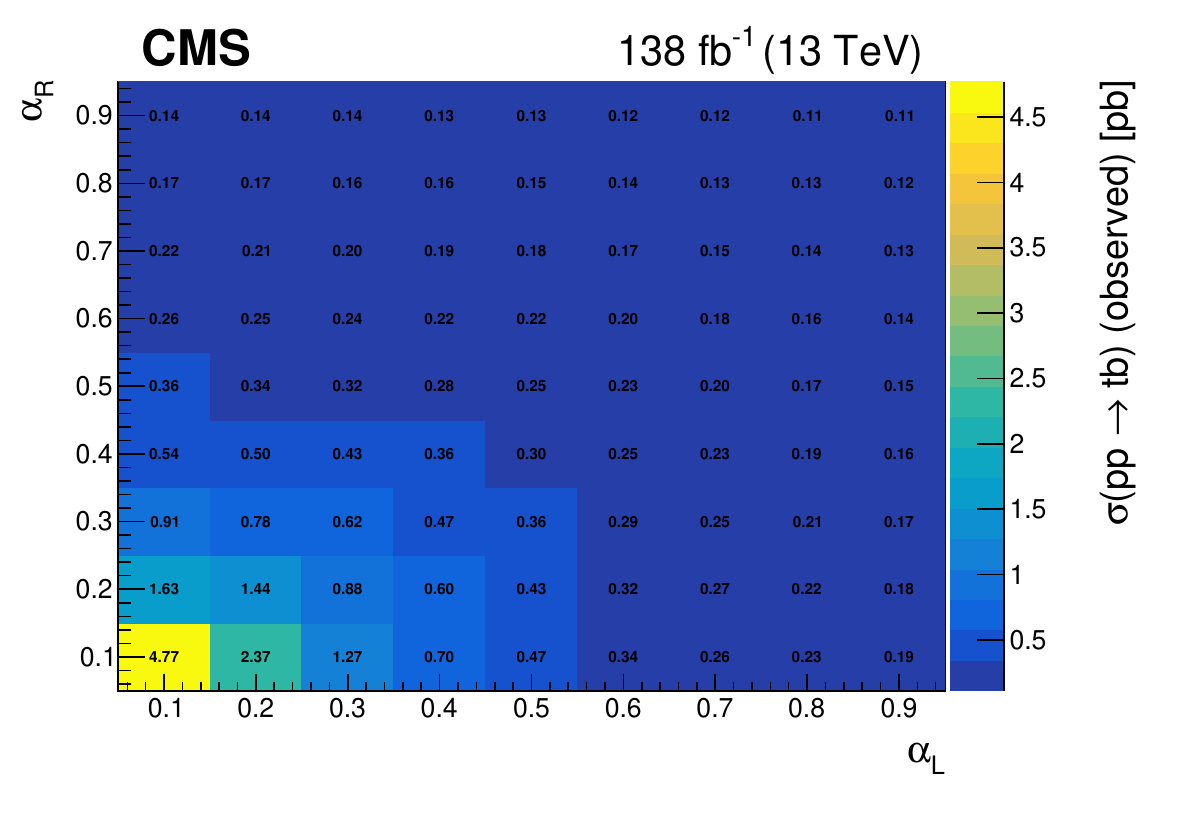}
\includegraphics[width=.48\textwidth]{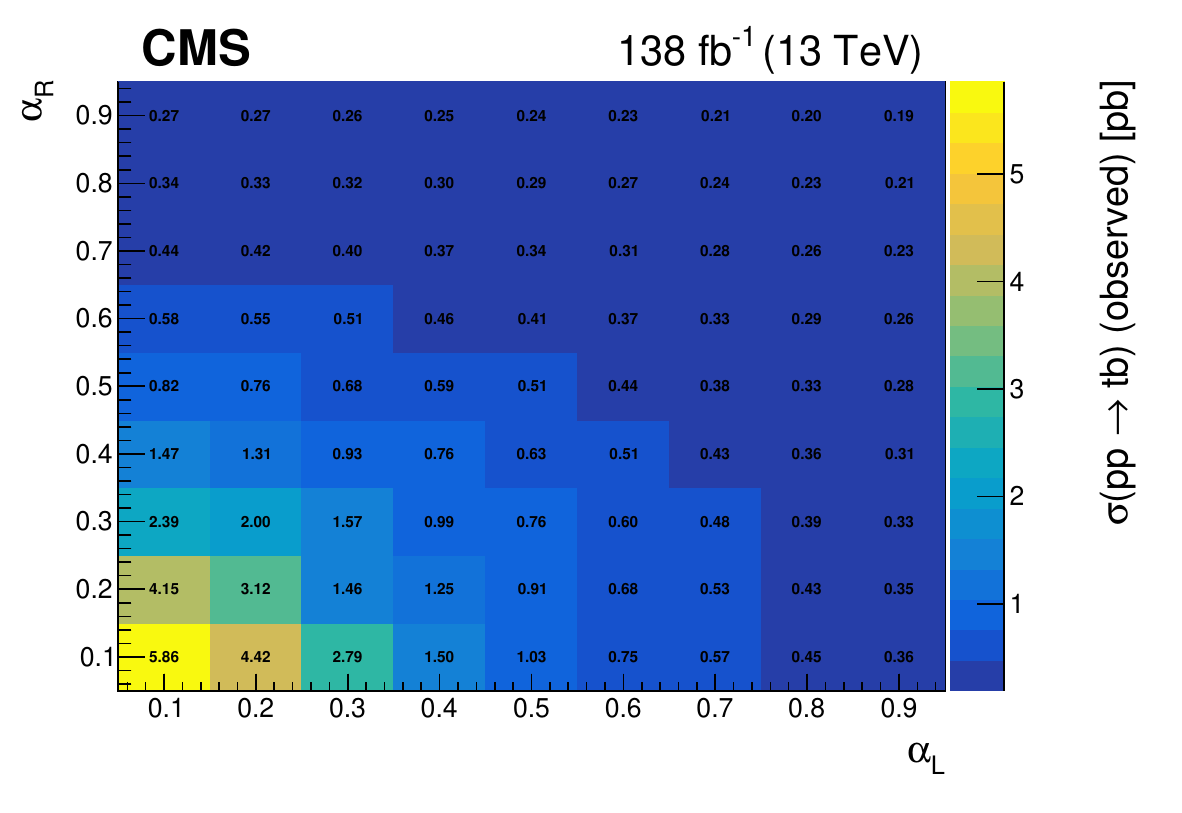}\\
\includegraphics[width=.48\textwidth]{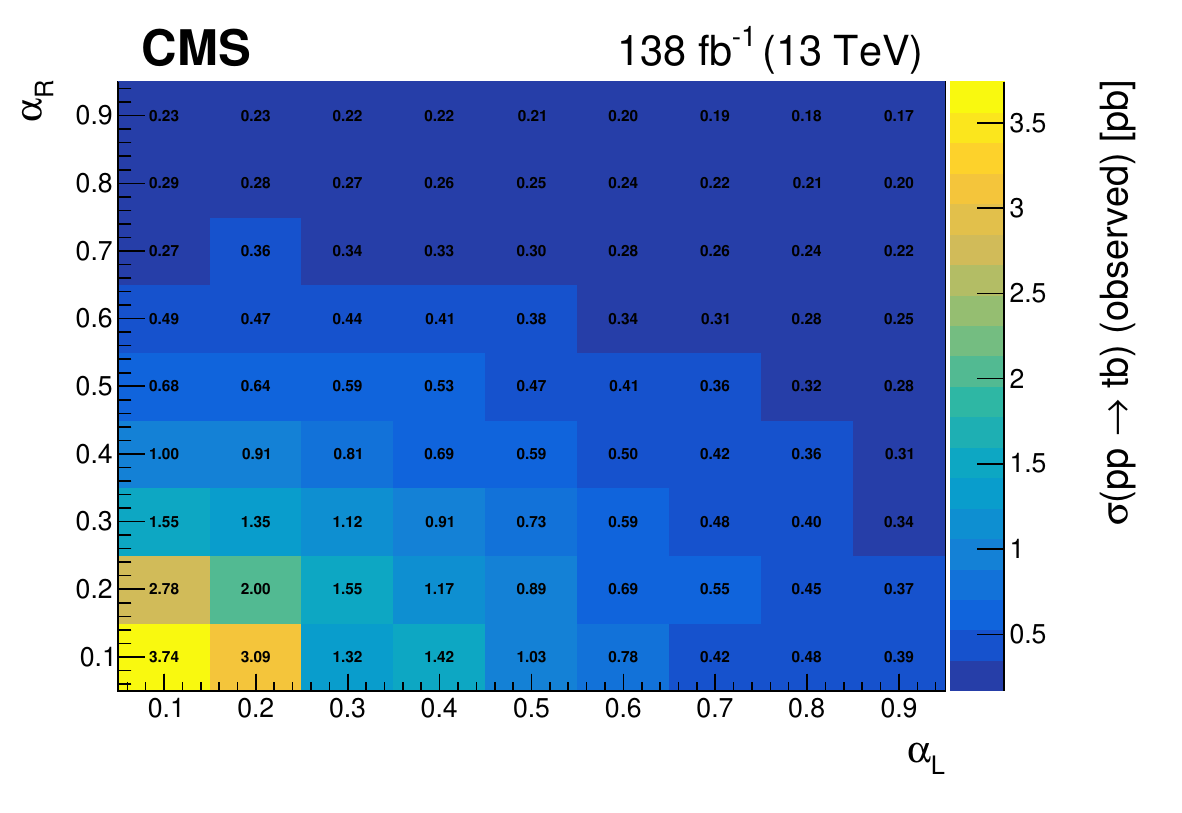}
\includegraphics[width=.48\textwidth]{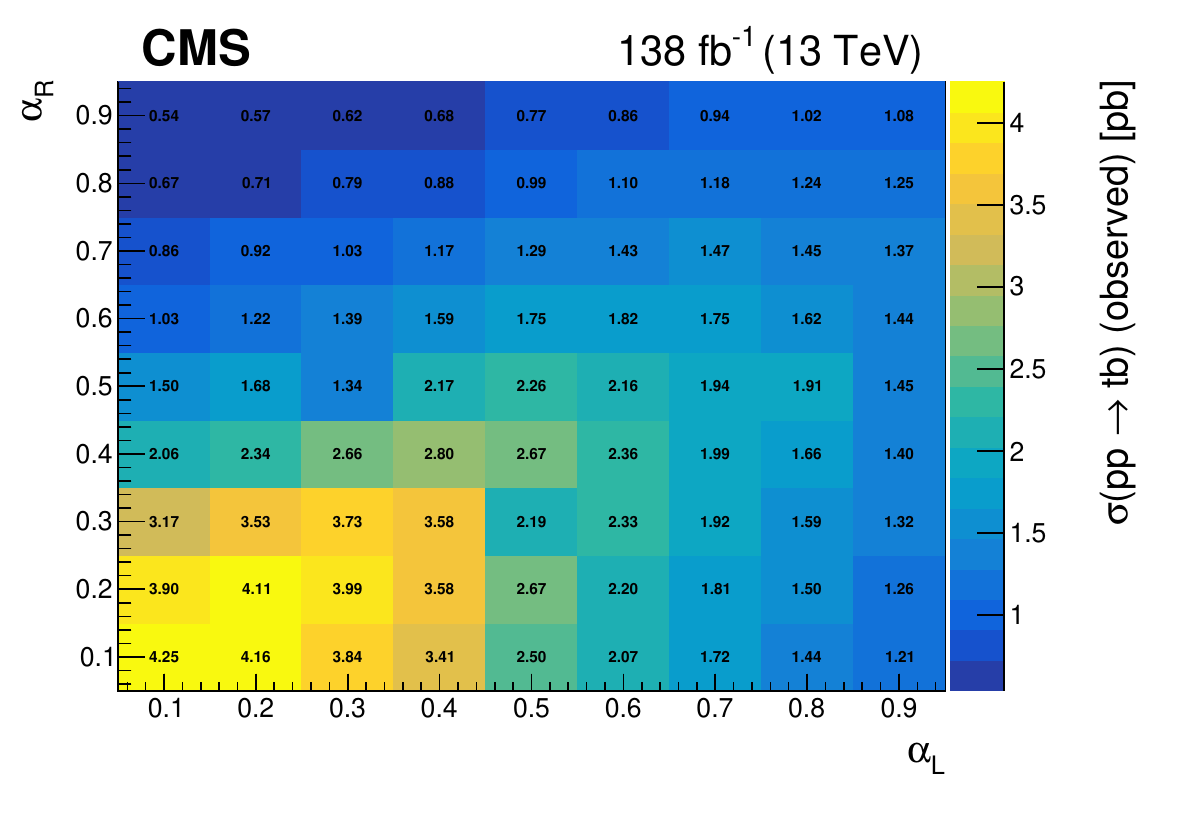}\\
\caption{Observed 95\% \CL upper limit on the production cross section for a generalized left-right coupling of the \PWpr boson to a \PQt and a \PQb quark for a mass of the \PWpr boson of 2\TeV (upper left), 2.8\TeV (upper right), 3.6\TeV (middle left), 4.4\TeV (middle right), 5.2\TeV (lower left), and of 6\TeV (lower right).}
\label{fig:xs2Dexcl}
\end{figure}

Making use of this formula, we can evaluate upper limits on the production cross section, to allow for re-interpretation in a wide array of models, or on the \PWpr boson once the theoretical cross section is specified.
The observed 95\% \CL upper limits on the production cross section for a generalized left-right coupling of the \PWpr boson to a \PQt and a \PQb quark are obtained for masses of the \PWpr boson between 2 and 6\TeV in steps of 0.8\TeV. The observed limits are reported in Fig.~\ref{fig:xs2Dexcl}.

\clearpage

\begin{figure}[htb]
\centering
\includegraphics[width=.48\textwidth]{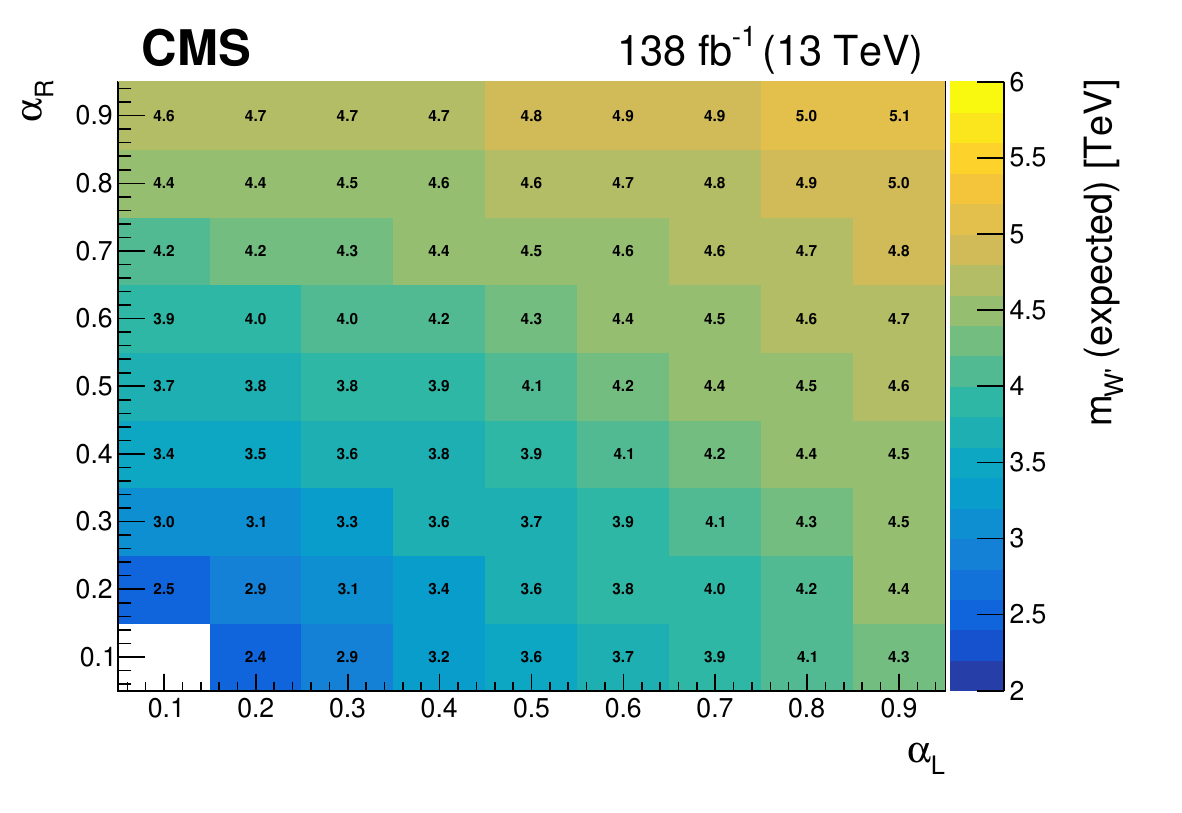}
\includegraphics[width=.48\textwidth]{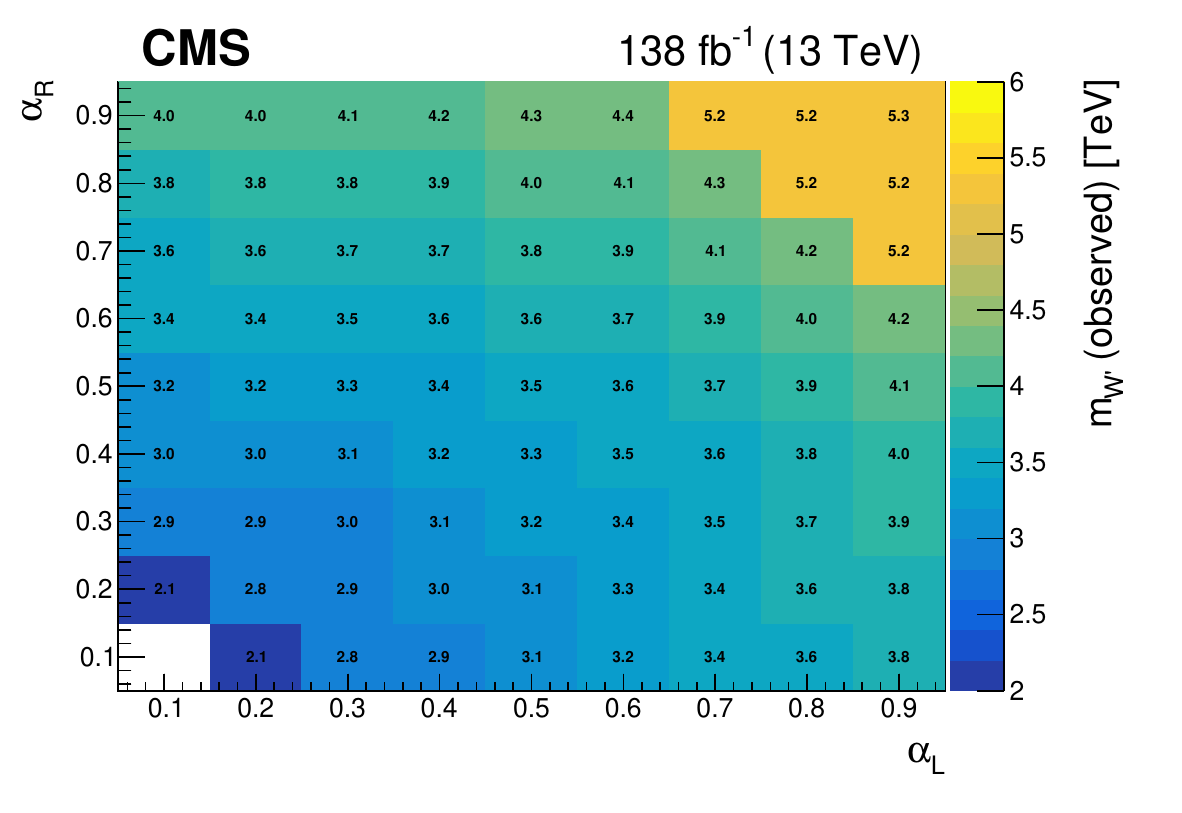}\\
    \caption{Expected (left) and observed (right) 95\% \CL lower limit on \mwp for a generalized left-right coupling of the \PWpr boson to a \PQt and a \PQb quark.}
\label{fig:mass2Dexcl}
\end{figure}

Limits on \mwp are also obtained for all tested values of the generalized left-right coupling of the \PWpr boson to a \PQt and a \PQb quark. Figure~\ref{fig:mass2Dexcl} shows the expected and observed limits on \mwp.

\section{Summary}
A search is presented for \PWpr bosons decaying to a top and a bottom quark in leptonic final states, making use of 138\fbinv of proton-proton collision data collected with the CMS detector at the LHC. Good agreement between data and the standard model expectation is observed. 

Upper limits at 95\% confidence level are set on the product of the \PWpr production cross section and the branching fraction of $\PWpr\to \PQt\PQb$. Multiple hypotheses are considered for the new particle mass, width, and chirality.
For a 1\% relative width hypothesis, purely right-handed \PWpr bosons are excluded with masses lower than 4.3\TeV. Production cross sections above 66 to 2\unit{fb} are excluded for masses between 2 and 6\TeV. Purely left-handed \PWpr bosons with a 1\% relative decay width are excluded for masses lower than 3.9\TeV.

The largest excess, with a local (global) significance of 2.6 (2.0) standard deviations, is observed for a hypothesized right-handed \PWpr boson with a mass of 3.8\TeV and a relative width of 1\%.

For a 10\% relative width hypothesis, purely right-handed \PWpr bosons are excluded with masses lower than 2.7\TeV. 
Purely left-handed \PWpr bosons are excluded with masses lower than 2.5\TeV for a relative width of 10\% of the \PWpr boson mass. Limits on the production cross section of a \PWpr boson with relative widths of 20\% and 30\% are set for purely left- and right-handed couplings.
Scenarios with the presence of both left- and right-handed couplings are also tested and limits on their production cross sections are set. For these scenarios, exclusion limits on the \PWpr boson mass are provided for the considered model. This is the first time an analysis probes both the width and the chirality of the \PWpr boson.

\begin{acknowledgments}
We congratulate our colleagues in the CERN accelerator departments for the excellent performance of the LHC and thank the technical and administrative staffs at CERN and at other CMS institutes for their contributions to the success of the CMS effort. In addition, we gratefully acknowledge the computing centers and personnel of the Worldwide LHC Computing Grid and other centers for delivering so effectively the computing infrastructure essential to our analyses. Finally, we acknowledge the enduring support for the construction and operation of the LHC, the CMS detector, and the supporting computing infrastructure provided by the following funding agencies: SC (Armenia), BMBWF and FWF (Austria); FNRS and FWO (Belgium); CNPq, CAPES, FAPERJ, FAPERGS, and FAPESP (Brazil); MES and BNSF (Bulgaria); CERN; CAS, MoST, and NSFC (China); MINCIENCIAS (Colombia); MSES and CSF (Croatia); RIF (Cyprus); SENESCYT (Ecuador); MoER, ERC PUT and ERDF (Estonia); Academy of Finland, MEC, and HIP (Finland); CEA and CNRS/IN2P3 (France); SRNSF (Georgia); BMBF, DFG, and HGF (Germany); GSRI (Greece); NKFIH (Hungary); DAE and DST (India); IPM (Iran); SFI (Ireland); INFN (Italy); MSIP and NRF (Republic of Korea); MES (Latvia); LAS (Lithuania); MOE and UM (Malaysia); BUAP, CINVESTAV, CONACYT, LNS, SEP, and UASLP-FAI (Mexico); MOS (Montenegro); MBIE (New Zealand); PAEC (Pakistan); MES and NSC (Poland); FCT (Portugal); MESTD (Serbia); MCIN/AEI and PCTI (Spain); MOSTR (Sri Lanka); Swiss Funding Agencies (Switzerland); MST (Taipei); MHESI and NSTDA (Thailand); TUBITAK and TENMAK (Turkey); NASU (Ukraine); STFC (United Kingdom); DOE and NSF (USA).

\hyphenation{Rachada-pisek} Individuals have received support from the Marie-Curie program and the European Research Council and Horizon 2020 Grant, contract Nos.\ 675440, 724704, 752730, 758316, 765710, 824093, and COST Action CA16108 (European Union); the Leventis Foundation; the Alfred P.\ Sloan Foundation; the Alexander von Humboldt Foundation; the Science Committee, project no. 22rl-037 (Armenia); the Belgian Federal Science Policy Office; the Fonds pour la Formation \`a la Recherche dans l'Industrie et dans l'Agriculture (FRIA-Belgium); the Agentschap voor Innovatie door Wetenschap en Technologie (IWT-Belgium); the F.R.S.-FNRS and FWO (Belgium) under the ``Excellence of Science -- EOS" -- be.h project n.\ 30820817; the Beijing Municipal Science \& Technology Commission, No. Z191100007219010 and Fundamental Research Funds for the Central Universities (China); the Ministry of Education, Youth and Sports (MEYS) of the Czech Republic; the Shota Rustaveli National Science Foundation, grant FR-22-985 (Georgia); the Deutsche Forschungsgemeinschaft (DFG), under Germany's Excellence Strategy -- EXC 2121 ``Quantum Universe" -- 390833306, and under project number 400140256 - GRK2497; the Hellenic Foundation for Research and Innovation (HFRI), Project Number 2288 (Greece); the Hungarian Academy of Sciences, the New National Excellence Program - \'UNKP, the NKFIH research grants K 124845, K 124850, K 128713, K 128786, K 129058, K 131991, K 133046, K 138136, K 143460, K 143477, 2020-2.2.1-ED-2021-00181, and TKP2021-NKTA-64 (Hungary); the Council of Science and Industrial Research, India; ICSC -- National Research Center for High Performance Computing, Big Data and Quantum Computing, funded by the NextGenerationEU program (Italy); the Latvian Council of Science; the Ministry of Education and Science, project no. 2022/WK/14, and the National Science Center, contracts Opus 2021/41/B/ST2/01369 and 2021/43/B/ST2/01552 (Poland); the Funda\c{c}\~ao para a Ci\^encia e a Tecnologia, grant CEECIND/01334/2018 (Portugal); the National Priorities Research Program by Qatar National Research Fund; MCIN/AEI/10.13039/501100011033, ERDF ``a way of making Europe", and the Programa Estatal de Fomento de la Investigaci{\'o}n Cient{\'i}fica y T{\'e}cnica de Excelencia Mar\'{\i}a de Maeztu, grant MDM-2017-0765 and Programa Severo Ochoa del Principado de Asturias (Spain); the Chulalongkorn Academic into Its 2nd Century Project Advancement Project, and the National Science, Research and Innovation Fund via the Program Management Unit for Human Resources \& Institutional Development, Research and Innovation, grant B05F650021 (Thailand); the Kavli Foundation; the Nvidia Corporation; the SuperMicro Corporation; the Welch Foundation, contract C-1845; and the Weston Havens Foundation (USA).
\end{acknowledgments}

\bibliography{auto_generated}
\clearpage
\numberwithin{table}{section}
\numberwithin{figure}{section}
\appendix

\section{Alternative cross sections}
\label{sec:alter_xs}
\begin{figure}[tbh]
\centering
\includegraphics[width=.48\textwidth]{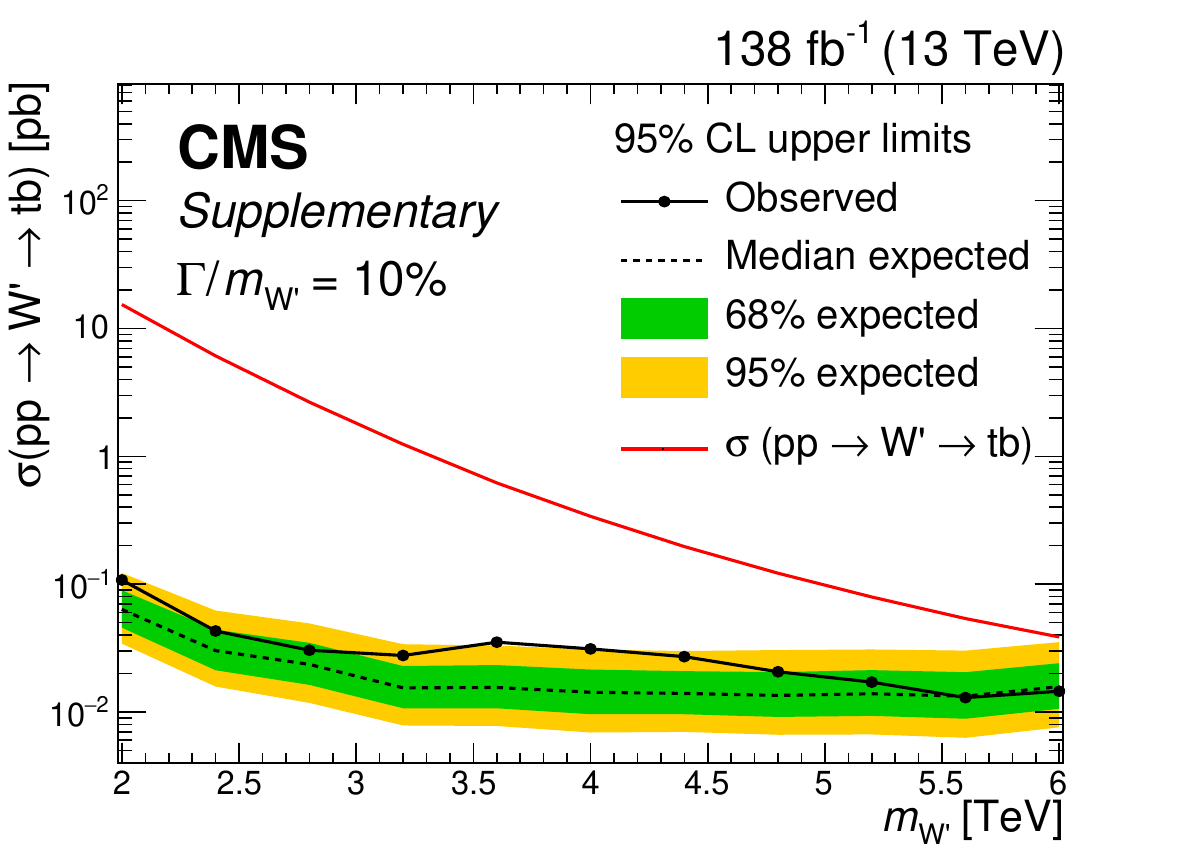}
\includegraphics[width=.48\textwidth]{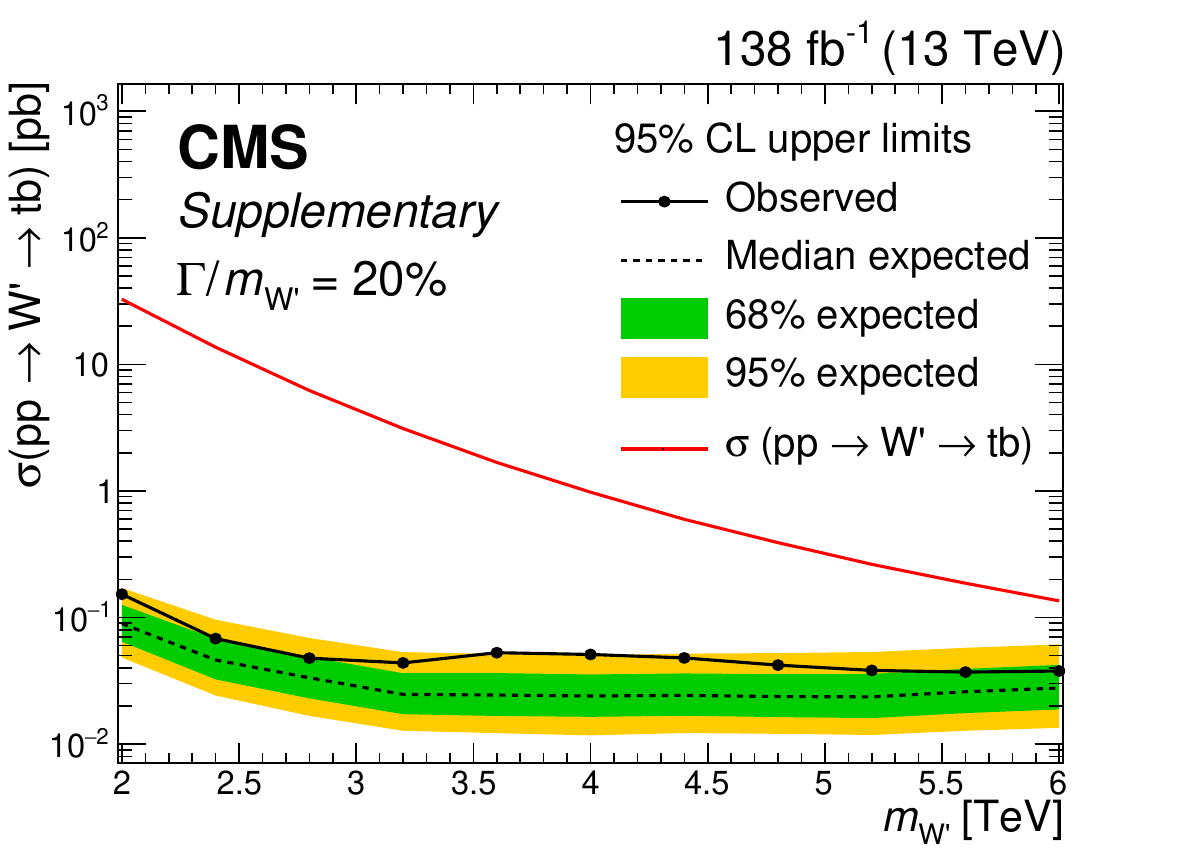}\\
\includegraphics[width=.48\textwidth]{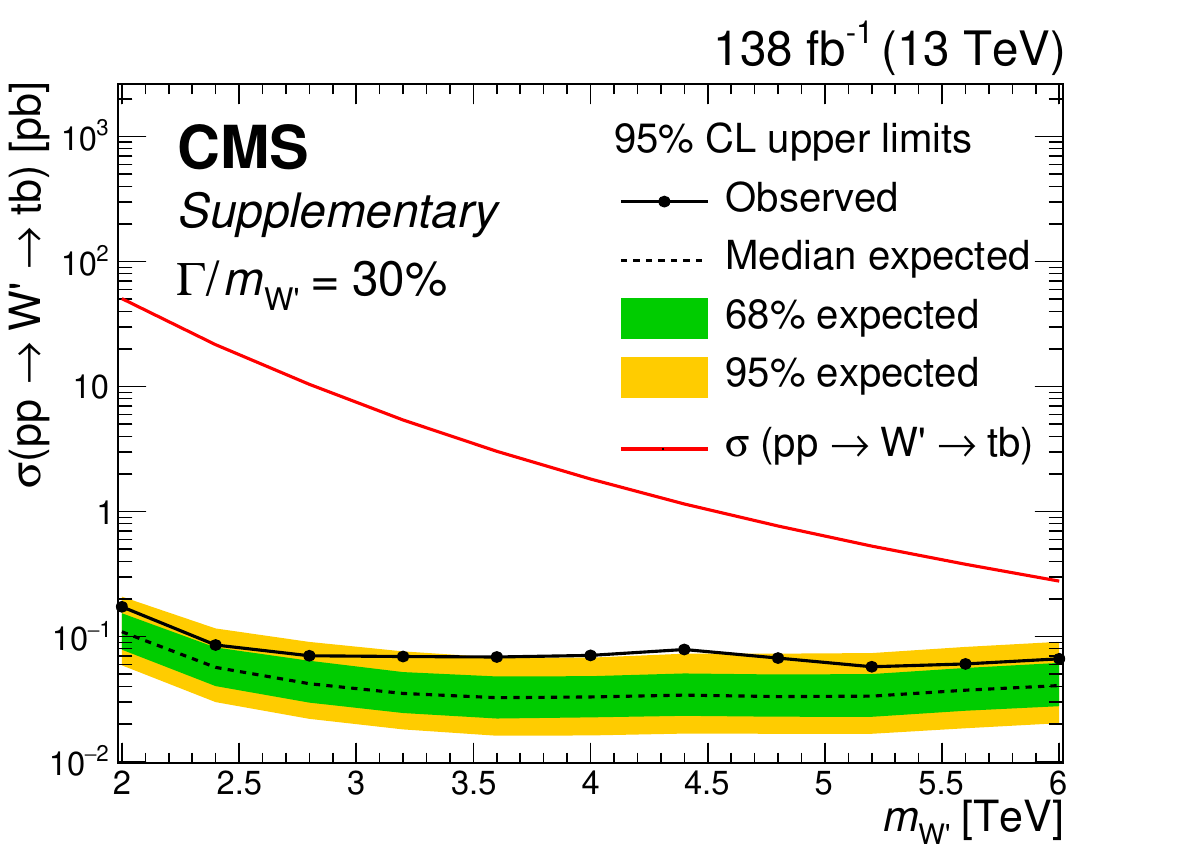}
    \caption{Observed and expected 95\% \CL upper limits on the product of the production cross section for a right-handed \PWpr boson and the $\PWpr \to \PQt\PQb$ branching fraction, as functions of the \mwp for a relative width of 10\% (upper left), 20\% (upper right), and 30\% (lower).
    The inner (green) band and the outer (yellow) band indicate the regions containing 68 and 95\%, respectively, of the distribution of limits expected under the background-only hypothesis.
    The solid curves show the theoretical expectation at LO in the case that the couplings, and thus the partial widths, are varied together with the total width. In this interpretation the branching fraction of the $\PWpr \to \PQt\PQb$ decays is the same for each value of the width of the \PWpr boson.}
\label{fig:alternative_sigma_mass_limits_RH}
\end{figure}

Upper limits are also provided in the case that the couplings, and thus the partial widths, are varied together with the total width. In this interpretation the branching fraction of the \PWpr $\to \PQt\PQb$ decays is the same for each value of the width of the \PWpr boson, where no additional decays are hypothesized, but the coupling strength is increased by a factor
$(\Gamma/\mwp)^{2}$. Figure~\ref{fig:alternative_sigma_mass_limits_RH} shows the limits for an RH \PWpr boson with these alternative theoretical cross sections. 

\cleardoublepage \section{The CMS Collaboration \label{app:collab}}\begin{sloppypar}\hyphenpenalty=5000\widowpenalty=500\clubpenalty=5000
\cmsinstitute{Yerevan Physics Institute, Yerevan, Armenia}
{\tolerance=6000
A.~Hayrapetyan, A.~Tumasyan\cmsAuthorMark{1}\cmsorcid{0009-0000-0684-6742}
\par}
\cmsinstitute{Institut f\"{u}r Hochenergiephysik, Vienna, Austria}
{\tolerance=6000
W.~Adam\cmsorcid{0000-0001-9099-4341}, J.W.~Andrejkovic, T.~Bergauer\cmsorcid{0000-0002-5786-0293}, S.~Chatterjee\cmsorcid{0000-0003-2660-0349}, K.~Damanakis\cmsorcid{0000-0001-5389-2872}, M.~Dragicevic\cmsorcid{0000-0003-1967-6783}, A.~Escalante~Del~Valle\cmsorcid{0000-0002-9702-6359}, P.S.~Hussain\cmsorcid{0000-0002-4825-5278}, M.~Jeitler\cmsAuthorMark{2}\cmsorcid{0000-0002-5141-9560}, N.~Krammer\cmsorcid{0000-0002-0548-0985}, D.~Liko\cmsorcid{0000-0002-3380-473X}, I.~Mikulec\cmsorcid{0000-0003-0385-2746}, J.~Schieck\cmsAuthorMark{2}\cmsorcid{0000-0002-1058-8093}, R.~Sch\"{o}fbeck\cmsorcid{0000-0002-2332-8784}, D.~Schwarz\cmsorcid{0000-0002-3821-7331}, M.~Sonawane\cmsorcid{0000-0003-0510-7010}, S.~Templ\cmsorcid{0000-0003-3137-5692}, W.~Waltenberger\cmsorcid{0000-0002-6215-7228}, C.-E.~Wulz\cmsAuthorMark{2}\cmsorcid{0000-0001-9226-5812}
% ...
\par}
\vskip\cmsinstskip
\dag:~Deceased\\
$^{1}$Also at Yerevan State University, Yerevan, Armenia\\
$^{2}$Also at TU Wien, Vienna, Austria\\
% ...
\end{sloppypar}
%%% END EDITABLE REGION %%%
% skeleton_end
\end{document}